\documentstyle[12pt,subeqnarray,psfig]{article}
\begin{document}
%
%
%                      --------------
%                      | versione B |
%                      -------------- 
%
%\lhead[\thepage]{R. Caimmi: R3 fluids}
%\rhead[Astron. Nachr./AN~{\bf XXX} (2005) X]{\thepage}
%\headnote{Astron. Nachr./AN {\bf 33X} {2005} X, XXX-XXX}
%
\newenvironment{lefteqnarray}{\arraycolsep=0pt\begin{eqnarray}}
{\end{eqnarray}\protect\aftergroup\ignorespaces}
\newenvironment{lefteqnarray*}{\arraycolsep=0pt\begin{eqnarray*}}
{\end{eqnarray*}\protect\aftergroup\ignorespaces}
\newenvironment{leftsubeqnarray}{\arraycolsep=0pt\begin{subeqnarray}}
{\end{subeqnarray}\protect\aftergroup\ignorespaces}
\newcommand{\displayfrac}[2]{\frac{\displaystyle #1}{\displaystyle #2}}
\newcommand{\diff}{{\rm\,d}}
\newcommand{\img}{{\rm i}}
\newcommand{\appleq}{\stackrel{<}{\sim}}
\newcommand{\appgeq}{\stackrel{>}{\sim}}
\newcommand{\Int}{\mathop{\rm Int}\nolimits}
\newcommand{\Nint}{\mathop{\rm Nint}\nolimits}
\newcommand{\Min}{\mathop{\rm min}\nolimits}
\newcommand{\Max}{\mathop{\rm max}\nolimits}
\newcommand{\Sgn}{\mathop{\rm Sgn}\nolimits}
%newcommand{\erf}{\mathop{\rm erf}\nolimits}
%\newcommand{\psfc}{\mathop{\rm psfc}\nolimits}
%\newcommand{\Psf}{\mathop{\rm psf}\nolimits}
\newcommand{\arcsinh}{\mathop{\rm arcsinh}\nolimits}
\newcommand{\vers}{\mathop{\overrightarrow{\rm vers}}\nolimits}

\title{R3 fluids}   % ted density profiles. II. \\
%    Anisotropy and rotation}

\author{{R. Caimmi}\footnote{
%\titlerunning{R3 fluids}
%
%\author{R. Caimmi}
%
{\it Dipartimento di Astronomia, Universit\`a di Padova,
              Vicolo Osservatorio 2, I-35122 Padova, Italy -} 
              email: caimmi@pd.astro.it}
\phantom{agga}}

%\date{Received..............................................$\qquad$ Accepted...................................}
\maketitle
\begin{quotation}
\section*{}
\begin{Large}
\begin{center}

Abstract

\end{center}
\end{Large}
\begin{small}

The current paper is aimed in getting more
insight on three main points concerning large-scale
astrophysical systems, namely: (i) formulation of
tensor virial equations from the standpoint of
analytical mechanics; (ii) investigation on the
role of systematic and random motions with respect
to virial equilibrium configurations; (iii) extent to
which systematic and random motions are equivalent
in flattening or elongating the shape of a mass
distribution.   The tensor virial equations are
formulated regardless from the nature of the
system and its constituents, by generalizing
and extending a procedure used for the scalar
virial equations, in presence of discrete
subunits (Landau \& Lifchitz 1966).
%, Chap.\,II,\S\,10).
In particular, the self potential-energy
tensor is shown to be symmetric with respect
to the exchange of the indices, $(E_{\rm pot})_{pq}=
(E_{\rm pot})_{qp}$.   Then the results are extended
to continuous mass distributions.   The role of
systematic and random motions in collisionless,
ideal, self-gravitating fluids, is analysed in
detail including radial and tangential velocity
dispersion on the equatorial plane, and the
related mean angular velocity,
$\overline{\Omega}$, is
conceived as a figure rotation.   R3 fluids
are defined as ideal, self-gravitating fluids
in virial equilibrium, with systematic rotation
around a principal axis of inertia.   The
related virial equations are written in terms
of the moment of inertia tensor, $I_{pq}$,
the self potential-energy tensor, $(E_{\rm pot})_
{pq}$, and the generalized anisotropy tensor,
$\zeta_{pq}$ (Caimmi \& Marmo 2005; Caimmi 
2006a).   Additional effort is devoted to
the investigation of the properties of
axisymmetric and triaxial configurations.
A unified theory of systematic and random
motions is developed for R3 fluids, taking
into consideration imaginary rotation 
(Caimmi 1996b, 2006a).   The effect of
random motion excess is shown to be 
equivalent to an additional real or
imaginary rotation, respectively, inducing
flattening (along the equatorial plane) or
elongation (along the rotation axis).
Then it is realized that a R3 fluid
always admits an adjoint configuration
with isotropic random velocity
distribution.  In addition, further 
constraints are established on the amount 
of random velocity anisotropy along the 
principal axes, for triaxial configurations.
A necessary condition is formulated for
the occurrence of bifurcation
points from axisymmetric to triaxial 
configurations in virial equilibrium,
which is independent of the anisotropy
parameters.   A particularization is made to
the special case of homeoidally striated
Jacobi ellipsoid, and some
previously known results (Caimmi 2006a)
are reproduced.

\noindent
{\it keywords - scalar virial theorem; tensor virial theorem;
collisionless, self-gravitating ideal fluids; random velocity
distributions; stellar systems.}
%   }

%   \maketitle
\end{small}
\end{quotation}
%
%________________________________________________________________

\section{Introduction}\label{intro}
Large-scale celestial objects, such as
stellar systems, galaxy clusters, and
(non baryonic) dark matter haloes predicted
by current $\Lambda$CDM cosmologies, may
safely be represented as collisionless,
ideal self-gravitating fluids.   The
related flow equation takes the same
formal expression as in their collisional
counterpart, with the exception that the
pressure force is generalized in terms of
a stress tensor, allowing different rms
velocities along different directions
[e.g., Binney \& Tremaine 1987 (hereafter
quoted as BT87), Chap.\,4, \S\,2].
Accordingly, collisionless fluids can
be flattened equally well by rotation
(with respect to a selected axis) and/or
anisotropic random velocity distribution
i.e. anisotropic pressure (e.g., Caimmi
2006a, hereafter quoted as C06%
\footnote{A more extended file including
an earlier version of the above quoted
paper is available at the arxiv electronic
site, as astro-ph/0507314.}).
In fact, giant elliptical galaxies exhibit 
a negligible amount of (systematic)
rotation, and their shape is mainly due
to anisotropic pressure (e.g., Bertola
\& Capaccioli 1975; Illingworth 1977,
1981; Schechter \& Gunn 1979; BT87, 
Chap.\,4, \S\,36).

Collisionless fluids of astrophysical
and cosmological interest range over 
about ten decades in mass, from globular 
clusters to galaxy clusters say, provided
gaseous i.e. collisional component may
safely be neglected.   Therefore, it 
seems necessary to investigate the role
of systematic and random motions in
making virialized collisionless fluids.
To this respect, the virial theorem in
tensor form may be a useful tool.
According to the standard procedure,
the tensor virial equations are 
determined along the following steps
(e.g., Binney 1978, 2005; Wiegandt
1982a,b; BT87, Chap.\,4, \S\,3): (a)
start with the collisionless Boltzmann
equations; (b) derive a set of moment 
equations; (c) integrate the above set
of moment equations, under some 
simplifying assunptions.
 
To the (limited) knowledge of the author,
no attempt can be found in the literature
where (i) the tensor virial equations are
formulated from the standpoint of analytical
mechanics; (ii) the role of systematic and
random motions is clearly stated, and (iii)
the equivalence between systematic and random
motions in flattening or elongating the
boundary, is clearly established.

Concerning (i), the tensor virial equations
could be determined regardless from the
nature of the system and its constituents,
by generalizing and extending a procedure
used for the scalar virial equations [Landau
\& Lifchitz 1966 (hereafter quoted as LL66),
Chap.\,II, \S\,10].
With regard to (ii), preliminary considerations
reported in previous attempts [Caimmi 1996a,b;
Caimmi \& Marmo 2005 (hereafter quoted as CM05);
C06] should be further improved and developed.
In dealing with (iii), the definition of imaginary
rotation allows an interpretation of rms velocity
excess in terms of systematic rotation around a
fixed principal axis (Caimmi 1996b; C06; Caimmi
2006b%
\footnote{A more extended file including
an earlier version of the above quoted paper is
available at the arxiv electronic site, as
astro-ph/0507314.})
which
could be inserted in the context under discussion.
A detailed investigation on the above mentioned
points makes the aim of the present attempt.

The current paper is organized as follows.   A 
general formulation of the tensor virial theorem
which holds, in particular, for the gravitational
interaction, is provided in Sect.\,\ref{tvt}.
The properties of R3 fluids, defined as ideal, 
self-gravitating fluids in virial equilibrium, 
rotating around a principal axis, are studied in 
Sect.\,\ref{syra}, and a unified theory of
systematic and random motions is provided in
Sect.\,\ref{imro}, where some general relations
are particularized to the special case of
homeoidally striated Jacobi ellipsoids, and
previously known results (Caimmi 1996a,b; C06)
are reproduced.   Some concluding remarks are 
drawn in Sect.\,\ref{conc}, and a few arguments 
are treated with more detail in the Appendix.   

\section{The tensor virial theorem}
\label{tvt}

A general procedure used for the formulation
of the scalar virial theorem (LL66, Chap.\,II,
\S\,10) shall be followed here in the derivation
of the tensor virial theorem.   Let us take
into consideration a mechanical system made
of $N$ particles, referred to an inertial
frame.   Let $(x_i)_r$, $(v_i)_r$, be the
position and velocity components related to
$i$ particle, and $m_i$ the mass, $1\le i
\le N$, $1\le r\le3$.

The kinetic-energy tensor:
\begin{equation}
\label{eq:Tpq}
(E_{\rm kin})_{pq}=\frac12\sum_{i=1}^Nm_i
(v_i)_p(v_i)_q~~;
\end{equation}
is a function of $2N$ or $N$ variables,
$(v_i)_r$, $1\le i\le N$, $1\le
p\le3$, $1\le q\le3$, for selected $p$
and $q$, according if the tensor
components are diagonal or non diagonal,
respectively.   The kinetic-energy tensor
is manifestly symmetric with respect to
the indices:
\begin{equation}
\label{eq:Tsim}
(E_{\rm kin})_{pq}=(E_{\rm kin})_{qp}~~;
\end{equation}
and the trace is the kinetic energy:
\begin{equation}
\label{eq:T}
E_{\rm kin}=\sum_{s=1}^3(E_{\rm kin})_{ss}=\frac12
\sum_{i=1}^N\sum_{s=1}^3m_i(v_i)_s^2~~;
\end{equation}
which is a function of $3N$ variables,
$(v_i)_r$, $1\le i\le N$, $1\le r\le3$.
The first partial derivatives are:
\begin{equation}
\label{eq:dT}
(p_i)_r=\frac{\partial E_{\rm kin}}{\partial 
(v_i)_r}=m_i(v_i)_r~~;
\end{equation}
where $(p_i)_r$, $1\le i\le N$, $1\le r\le3$, 
is the impulse component of $i$ particle (e.g.,
LL66, Chap.\,II, \S\,7).

The combination of Eqs.\,(\ref{eq:Tpq}) and
(\ref{eq:dT}) yields:
\begin{equation}
\label{eq:Tpq1}
2(E_{\rm kin})_{pq}=\sum_{i=1}^N(v_i)_p
(p_i)_q~~;
\end{equation}
which is equivalent to:
\begin{equation}
\label{eq:Tpq2}
2(E_{\rm kin})_{pq}=\frac{\diff}{\diff t}\left[
\sum_{i=1}^N(x_i)_p(p_i)_q\right]-\sum_{i=1}^N
(x_i)_p(\dot{p}_i)_q~~;
\end{equation}
or, using Newton's equations (e.g., LL66,
Cap.\,I, \S\,5):
\begin{equation}
\label{eq:Tpq3}
2(E_{\rm kin})_{pq}=\frac{\diff}{\diff t}\left[
\sum_{i=1}^N(x_i)_p(p_i)_q\right]+\sum_{i=1}^N
(x_i)_p\frac{\partial E_{\rm pot}}{\partial(x_i)_q}~~;
\end{equation}
where $E_{\rm pot}[(x_i)_r]$, $1\le i\le N$, $1\le r
\le3$, is the self potential energy.

The last term on the right-hand side of
Eq.\,(\ref{eq:Tpq3}) defines a tensor,
the trace of which is usually named the
virial of the system (Clausius 1870).
In the author's opinion, it would be
better to quote the virial and
its parent tensor as the virial potential
energy and the virial potential-energy
tensor, respectively.

If the self potential energy is a homogeneous
function of the coordinates, of degree
$\chi$, then the following relation holds:
\begin{equation}
\label{eq:Vhan}
E_{\rm pot}[\zeta(x_i)_r]=\zeta^\chi E_{\rm pot}
[(x_i)_r]~~;
\end{equation}
which, in turn, implies:
\begin{equation}
\label{eq:Vpqh}
(E_{\rm pot})_{pq}[\zeta(x_i)_r]=\zeta^\chi
(E_{\rm pot})_{pq}[(x_i)_r]~~;
\end{equation}
where $(E_{\rm pot})_{pq}$ is defined as the
self potential-energy tensor:
\begin{equation}
\label{eq:Vpq}
(E_{\rm pot})_{pq}=\frac1\chi\sum_{i=1}^N
(x_i)_p\frac{\partial E_{\rm pot}}{\partial
(x_i)_q}~~;
\end{equation}
and $\zeta$ is a generic real number,
provided $\zeta(x_i)_r$, $1\le i\le N$, 
$1\le r\le3$, belonges to the domain of
$E_{\rm pot}$.

With regard to the self potential energy, 
the Euler theorem reads:
\begin{equation}
\label{eq:VEul}
\sum_{s=1}^3\sum_{i=1}^N(x_i)_s\frac
{\partial E_{\rm pot}}{\partial(x_i)_s}=
\chi E_{\rm pot}~~;
\end{equation}
and the combination of Eqs.\,(\ref{eq:Vpq})
and (\ref{eq:VEul}) yields:
\begin{equation}
\label{eq:TrVpq}
\sum_{s=1}^3(E_{\rm pot})_{ss}=E_{\rm pot}~~;
\end{equation}
as expected.

The substitution of Eq.\,(\ref{eq:Vpq})
into (\ref{eq:Tpq3}) yields:
\begin{equation}
\label{eq:virg}
\frac{\diff}{\diff t}\left[\sum_{i=1}^N
(x_i)_p(p_i)_q\right]=2(E_{\rm kin})_{pq}-\chi
(E_{\rm pot})_{pq}~~;
\end{equation}
and the sum of Eq.\,(\ref{eq:virg}) with
its counterpart where the indices, $p$
and $q$, are interchanged, reads:
\begin{equation}
\label{eq:virs}
\frac{\diff}{\diff t}\sum_{i=1}^N\left[
(x_i)_p(p_i)_q+(x_i)_q(p_i)_p\right]=2
[(E_{\rm kin})_{pq}+(E_{\rm kin})_{qp}]-\chi
[(E_{\rm pot})_{pq}+(E_{\rm pot})_{qp}]~~.
\end{equation}

Let us define the moment of inertia tensor%
\footnote{In this formulation, the moment
of inertia with respect to a coordinate
axis, $x_r$, is $I_r=I_{pp}+I_{qq}$, $r\ne
p\ne q$.   For a different formulation
where $I_r=I_{rr}$, $r=1,2,3,$ see LL66
(Chap.\,VI, \S\,32).}
[e.g., Chandrasekhar 1969 (hereafter
quoted as C69), Chap.\,2, \S\,9; BT87,
Chap.\,4, \S\,3]:
\begin{leftsubeqnarray}
\slabel{eq:Ipqa}
&& I_{pq}=\sum_{i=1}^Nm_i(x_i)_p(x_i)_q~~; \\
\slabel{eq:Ipqb}
&& \sum_{s=1}^3I_{ss}=I~~;
\label{seq:Ipq}
\end{leftsubeqnarray}
where $I$ is the total moment of
inertia of the system, with respect
to the centre of mass.   Owing to
Eq.\,(\ref{eq:dT}), the first
temporal derivative is:
\begin{equation}
\label{eq:dIpq}
\dot{I}_{pq}=\frac{\diff I_{pq}}{\diff t}=
\sum_{i=1}^N\left[(x_i)_p(p_i)_q+
(x_i)_q(p_i)_p\right]~~;
\end{equation}
and the combination of Eqs.\,(\ref{eq:Tsim}),
(\ref{eq:virs}), and (\ref{eq:dIpq}), yields:
\begin{equation}
\label{eq:virI}
\ddot{I}_{pq}=4(E_{\rm kin})_{pq}-\chi[(E_{\rm pot})_{pq}+
(E_{\rm pot})_{qp}]~~;
\end{equation}
on the other hand, the difference of 
Eq.\,(\ref{eq:virg}) with its counterpart 
where the indices, $p$ and $q$, are 
interchanged, reads:
\begin{equation}
\label{eq:vird}
\frac{\diff}{\diff t}\sum_{i=1}^N\left[
(x_i)_p(p_i)_q-(x_i)_q(p_i)_p\right]=-
\chi[(E_{\rm pot})_{pq}-(E_{\rm pot})_{qp}]~~.
\end{equation}
owing to Eq.\,(\ref{eq:Tsim}).

With regard to the vectors, $\vec{r}_i
[(x_i)_1,(x_i)_2,(x_i)_3]$ and $\vec{p}_i
[(p_i)_1,(p_i)_2,(p_i)_3]$, and to the
vector product, $\vec{J}_i=\vec{r}_i\times
\vec{p}_i$, the sum on the left-hand side
of Eq.\,(\ref{eq:vird}) reads (e.g.,
Spiegel 1968, Chap.\,2.2, \S\S\,11-12):
\begin{lefteqnarray}
\label{eq:Jr}
&& \sum_{i=1}^N\left[(x_i)_p(p_i)_q-(x_i)_q(p_i)_p\right]=
\sum_{i=1}^N\vers(x_r)\cdot(\vec{r}_i\times\vec{p}_i)
\nonumber \\
&& \quad=\vers(x_r)\cdot\sum_{i=1}^N\vec{J_i}=
\vers(x_r)\cdot\vec{J}=J_r~~;
\end{lefteqnarray}
where $\vers(x_r)$ is the versor, or unit vector,
parallel to the coordinate axis, $x_r$, $r\ne p
\ne q$, and $J$ is the total angular moment
of the system.

The combination of Eqs.\,(\ref{eq:vird})
and (\ref{eq:Jr}) yields:
\begin{equation}
\label{eq:dJr}
\frac{\diff J_r}{\diff t}=-\chi[(E_{\rm pot})_{pq}-
(E_{\rm pot})_{qp}]~~;
\end{equation}
and the conservation of angular momentum,
which always holds for isolated systems
(e.g., LL66, Chap.\,2, \S\,9), implies
the symmetry of the self potential-energy
tensor with respect to the exchange of
the indices:
\begin{equation}
\label{eq:Vpqs}
(E_{\rm pot})_{pq}=(E_{\rm pot})_{qp}~~;
\end{equation}
and Eq.\,(\ref{eq:virI}) takes the form:
\begin{equation}
\label{eq:virdI}
\frac12\ddot{I}_{pq}=2(E_{\rm kin})_{pq}-
\chi(E_{\rm pot})_{pq}~~;
\end{equation}
which makes the virial equations of the
second order (for the special case of
gravitational interaction, $\chi=-1$,
see e.g., C69, Chap.\,2, \S\,11; BT87,
Chap.\,4, \S\,3).

The further restriction:
\begin{equation}
\label{eq:dI20}
\ddot{I}_{pq}=0~~;\qquad
1\le p\le3~~;\qquad1\le q\le3~~;
\end{equation}
makes Eqs.\,(\ref{eq:virdI}) reduce to:
\begin{equation}
\label{eq:vire0}
2(E_{\rm kin})_{pq}-\chi(E_{\rm pot})_{pq}=0~~;\qquad
1\le p\le3~~;\qquad1\le q\le3~~;
\end{equation}
which is the expression of the virial
theorem in tensor form%
\footnote{Some authors prefer a more general
formulation, expressed by Eqs.\,(\ref{eq:virdI})
(e.g., BT87, Chap.\,4, \S\,3).   On the other hand,
a more restricted formulation, expressed by 
Eqs.\,(\ref{eq:vire0}), has a closer connection
with the scalar virial theorem, which explains
the choice adopted here.}.
Strictly speaking, it holds when the moment of
inertia tensor has a linear dependence on time,
$I_{pq}=k_{pq}t$, where $k_{pq}$ are constants.
The special case, $k_{pq}=0$, $1\le p\le3$, $1
\le q\le3$, is related to dynamical or hydrostatic
equilibrium (e.g., BT87, Chap.\,4, \S\,3).

An alternative restriction is that the first
time derivatives of the moment of inertia
tensor are bounded, as:
\begin{equation}
\label{eq:Ilim}
\vert\dot{I}_{pq}(t)\vert\le M_{pq}~~;\qquad
1\le p\le3~~;\qquad1\le q\le3~~;
\end{equation}
where $M_{pq}$ are convenient real numbers.
Accordingly, it can be seen that the time
average of the second time derivetives of
the moment of inertia tensor are null (e.g.,
LL66, Chap.\,II, \S\,10):
\begin{equation}
\label{eq:d2I0}
\overline{\ddot{I}_{pq}}=0~~;\qquad
1\le p\le3~~;\qquad1\le q\le3~~;
\end{equation}
which makes Eqs.\,(\ref{eq:virdI}) reduce to:
\begin{equation}
\label{eq:virm0}
2\overline{(E_{\rm kin})_{pq}}-\chi\overline
{(E_{\rm pot})_{pq}}=0~~;\qquad
1\le p\le3~~;\qquad1\le q\le3~~;
\end{equation}
where time averages are calculated over a
sufficiently long (ideally infinite) period
(e.g., LL66, Chap.\,2, \S\,10).   In presence 
of periodic motions (e.g.,
a homogeneous sphere undergoing coherent
oscillations), time averages can be 
calculated over a single (or a multiple)
period.

For sake of simplicity, in the following
the tensor virial theorem shall be expressed
by Eqs.\,(\ref{eq:vire0}) where the
kinetic-energy and self potential-energy
tensors are to be intended as instantaneous
or time averaged, according if the restriction
defined by Eq.\,(\ref{eq:dI20}) or (\ref{eq:d2I0})
holds.

The particularization of Eqs.\,(\ref{eq:vire0})
to diagonal components, after summation on both
sides, produces:
\begin{equation}
\label{eq:virte}
2E_{\rm kin}-\chi E_{\rm pot}=0~~;
\end{equation}
which is the expression of the virial
theorem in scalar form (e.g., LL66,
Chap.\,II, \S\,10).   Special cases
are (a) Newtonian and Coulombian
interaction, $\chi=-1$, and (b)
Hookeian interaction, $\chi=2$.
If the system is in dynamical or
hydrostatic equilibrium, mean values
coincide with instantaneous values.
%, and Eqs.\,(\ref{eq:vire0}) and (\ref
%{eq:virte}) read: $2(E_{\rm kin})_{pq}=
%\chi(E_{\rm pot})_{pq}$; $2(E_{\rm kin})=
%\chi(E_{\rm pot})$; respectively.

The above results are quite general
and hold regardless from the nature of
the system and its constituents,
provided no dissipation and/or
external interaction occur.   With
regard to a specified system, the
sole restrictions to be made are
(i) the evolution takes place within
a finite region of the phase hyperspace
i.e. $0\le\vert(x_i)_r\vert<M_{x_i}$,
$0\le\vert(v_i)_r\vert<M_{v_i}$, $1\le
i\le N$, $1\le r\le3$, where $M_{x_i}$,
$M_{v_i}$, are convenient real numbers,
and (ii) the
self potential energy is a homogeneous
function of the $3N$ coordinates, of
degree $\chi$.

In dealing with continuous matter
distributions instead of mass points,
the particle mass, $m_i$, has to be
replaced by the mass within an
infinitesimal volume element, $\diff
m=\rho(x_1,x_2,x_3)\diff x_1\diff x_2
\diff x_3$, where $\rho$ is the density,
and an integration has to be performed
over the whole volume, instead of a
summation on the coordinates related
to all the particles.   For further
details see e.g., Limber (1959).
Accordingly, the kinetic-energy tensor
and the kinetic energy attain their
usual expressions (C69, Chap.\,2,
\S\,9):
\begin{lefteqnarray}
\label{eq:Tpqc}
&&(E_{\rm kin})_{pq}=\frac12\int_S\rho(x_1,
x_2,x_3)v_pv_q\diff^3S~~; \\
\label{eq:Tc}
&&(E_{\rm kin})=\frac12\int_S\rho(x_1,
x_2,x_3)\sum_{s=1}^3v_s^2\diff^3S~~;
\end{lefteqnarray}
on the other hand, the self
potential-energy tensor and the self
potential energy read:
\begin{lefteqnarray}
\label{eq:Vpqc}
&&(E_{\rm pot})_{pq}=-\frac1\chi\int_S\rho_
\chi(x_1,x_2,x_3)x_p\frac{\partial{\cal V}}
{\partial x_q}\diff^3S~~; \\
\label{eq:Vc}
&&(E_{\rm pot})=-\frac1\chi\int_S\rho_\chi(x_1,
x_2,x_3)\sum_{s=1}^3x_s\frac{\partial{\cal V}}
{\partial x_s}\diff^3S~~;
\end{lefteqnarray}
where $\rho_\chi$ is a charge density,
${\cal V}$ is a potential function,
defined as the tidal potential energy
acting on the unit charge (related to
the interaction), placed at the point
under consideration.   For further details,
see Appendix A.

Let us define the total-energy tensor, as:
\begin{equation}
\label{eq:Epq}
E_{pq}=(E_{\rm kin})_{pq}+(E_{\rm pot})_{pq}~~;
\end{equation}
owing to Eqs.\,(\ref{eq:T}) and (\ref{eq:TrVpq}),
the related trace:
\begin{equation}
\label{eq:E}
E=\sum_{s=1}^3E_{ss}=E_{\rm kin}+E_{\rm pot}~~;
\end{equation}
is the total energy.

The combination of Eqs.\,(\ref{eq:vire0}),
(\ref{eq:Epq}), and (\ref{eq:virte}),
(\ref{eq:E}), respectively, yields:
\begin{lefteqnarray}
\label{eq:TEpq}
&&(E_{\rm kin})_{pq}=\frac\chi{\chi+2}E_{pq}~~; \\
\label{eq:VEpq}
&&(E_{\rm pot})_{pq}=\frac2{\chi+2}E_{pq}~~;
\end{lefteqnarray}
for tensor components, and:
\begin{lefteqnarray}
\label{eq:TE}
&&E_{\rm kin}=\frac\chi{\chi+2}E~~; \\
\label{eq:VE}
&&E_{\rm pot}=\frac2{\chi+2}E~~;
\end{lefteqnarray}
for tensor traces.

\section{Systematic and random motions}
\label{syra}
\subsection{Basic ideas}\label{baid}

Let a collisionless, self-gravitating fluid
be referred to an inertial frame, $({\sf O}
x_1x_2x_3)$, where (without loss of generality)
the origin coincides with the centre of mass.
The number of particles within an infinitesimal
hypervolume of the phase hyperspace at the time,
$t$, is:
\begin{equation}
\label{eq:d6N}
\diff^6{\cal N}=f(x_1,x_2,x_3,v_1,v_2,v_3,t)
\diff x_1\diff x_2\diff x_3\diff v_1\diff v_2
\diff v_3~~;
\end{equation}
where $f\ge0$ is the distribution function.
The number of particles within an infinitesimal
volume of the ordinary space at the time, $t$,
is:
\begin{equation}
\label{eq:d3N}
\diff^3{\cal N}=\diff x_1\diff x_2\diff x_3
\int\int\int f(x_1,x_2,x_3,v_1,v_2,v_3,t)
\diff v_1\diff v_2\diff v_3~~;
\end{equation}
where the integration has to be performed
over the whole volume in velocity space.
The number density related to the
infinitesimal volume element, $\diff^3S=
\diff x_1\diff x_2\diff x_3$, at the time,
$t$, is:
\begin{equation}
\label{eq:n}
n(x_1,x_2,x_3,t)=\frac{\diff^3{\cal N}}{\diff^3S}=
\int\int\int f(x_1,x_2,x_3,v_1,v_2,v_3,t)
\diff v_1\diff v_2\diff v_3~~;
\end{equation}
if, in addition, the total particle number,
${\cal N}$, and the total mass, $M$, are
conserved, then the following normalization
conditions hold:
\begin{lefteqnarray}
\label{eq:N}
&& \int\int\int\int\int\int f(x_1,x_2,x_3,v_1,v_2,v_3,t)
\diff x_1\diff x_2\diff x_3\diff v_1\diff v_2\diff v_3=
{\cal N}~~; \\
\label{eq:M}
&& \int\int\int\rho(x_1,x_2,x_3,t)\diff x_1\diff x_2
\diff x_3=M~~;
\end{lefteqnarray}
where $\rho$ is the mass density of the infinitesimal
volume element, $\diff^3S$, and the integrations have
to be carried over the whole hypervolume in phase
hyperspace and the whole volume in ordinary space,
respectively.

From a physical point of view, the volume element is
finite instead of infinitesimal, but still containing
a large amount of particles which, on the other hand,
is negligible with respect to the total number.
Accordingly, the following relations hold:
\begin{equation}
\label{eq:DMDN}
1\ll\Delta{\cal N}(x_1,x_2,x_3,t)\ll{\cal N}~~;
\qquad\Max(m_i)\ll\Delta M(x_1,x_2,x_3,t)\ll M~~;
\end{equation}
where $\Delta{\cal N}$ and $\Delta M$ represent
the particle total number and total mass within
the volume element, $\diff^3S$, at the time, $t$,
and $m_i$ is the mass of $i$ particle, $1\le i\le
\Delta{\cal N}$.   The related total mass,
$\Delta M$, may be expressed as:
\begin{equation}
\label{eq:DM}
\Delta M(x_1,x_2,x_3,t)=\sum_{i=1}^{\Delta{\cal N}}
m_i~~;
\end{equation}
and the mean particle mass within the volume element, 
$\diff^3S$, at the time, $t$, reads:
\begin{equation}
\label{eq:mmd}
\overline{m}(x_1,x_2,x_3,t)=\frac{\Delta M(x_1,x_2,x_3,t)}
{\Delta{\cal N}(x_1,x_2,x_3,t)}~~;
\end{equation}
according to the general definition of arithmetic
mean.

From the standpoint of a continuous mass distribution,
the following changes have to be made: ${\Delta{\cal N}
(x_1,x_2,x_3,t)}\to\diff^3{\cal N}$; $\Delta M(x_1,x_2,
x_3,t)\to\rho(x_1,x_2,x_3,t)\diff^3S$; and Eq.\,(\ref
{eq:mmd}) takes the form:
\begin{equation}
\label{eq:mmc}
\overline{m}(x_1,x_2,x_3,t)=\rho(x_1,x_2,x_3,t)\frac
{\diff^3S}{\diff^3{\cal N}}=\frac{\rho(x_1,x_2,x_3,t)}
{n(x_1,x_2,x_3,t)}~~;
\end{equation}
in terms of mass density and number density.

With regard to Eqs.\,(\ref{eq:N}) and (\ref{eq:M}),
equivalent expressions are:
\begin{lefteqnarray}
\label{eq:Nn}
&& \int\int\int n(x_1,x_2,x_3,t)
\diff x_1\diff x_2\diff x_3={\cal N}~~; \\
\label{eq:Mm}
&& \int\int\int\rho(x_1,x_2,x_3,t)\diff x_1\diff x_2
\diff x_3 \nonumber \\
&& \quad=\int\int\int\overline{m}(x_1,x_2,x_3,t)
n(x_1,x_2,x_3,t)\diff x_1\diff x_2\diff x_3=M~~;
\end{lefteqnarray}
and the division of both sides of Eq.\,(\ref{eq:Mm})
by their counterparts in Eq.\,(\ref{eq:Nn}), yields:
\begin{equation}
\label{eq:mmt}
\overline{m}=\frac M{{\cal N}}~~;
\end{equation}
where, owing to the theorem of the mean,
$\overline{m}$ is the
particle mass averaged over the whole volume.
Total mass and particle number conservation
imply a time independent mean particle mass,
$\overline{m}$.   If, in addition, particles
with different masses are uniformly distributed
throughout the whole volume, then the mean
particle mass within a generic volume element,
$\diff^3S$, equals the mean particle mass
within the boundary, as:
\begin{equation}
\label{eq:mtm}
\overline{m}(x_1,x_2,x_3,t)=\overline{m}~~;
\end{equation}
and the system may be considered, in any respect,
as made of ${\cal N}$ identical particles of mass
$\overline{m}$.   Accordingly, Eq.\,(\ref{eq:mmc})
reads:
\begin{equation}
\label{eq:dmmn}
\rho(x_1,x_2,x_3,t)=\overline{m}~n(x_1,x_2,x_3,t)~~;
\end{equation}
which implies direct proportionality between mass
density and number density.

Using again the theorem of the mean, let us define
the mean velocity component, $\overline{v_p}$, and
the mean product velocity component, $\overline{v_
pv_q}$, within a generic infinitesimal volume
element, $\diff^3S=\diff x_1\diff x_2\diff x_3$,
at the time, $t$, as:
\begin{lefteqnarray}
\label{eq:vpm}
&& \overline{v_p}(x_1,x_2,x_3,t)=\frac{
%\diff x_1\diff x_2\diff x_3
\int\int\int f(x_1,x_2,x_3,v_1,v_2,v_3,t)v_p\diff 
v_1\diff v_2\diff v_3}{
%\diff x_1\diff x_2\diff x_3
\int\int\int f(x_1,x_2,x_3,v_1,v_2,v_3,t)\diff v_1
\diff v_2\diff v_3}~~; \\
\label{eq:vpqm}
&& \overline{v_pv_q}(x_1,x_2,x_3,t)=\frac{
%\diff x_1\diff x_2\diff x_3
\int\int\int f(x_1,x_2,x_3,v_1,
v_2,v_3,t)v_pv_q\diff v_1\diff v_2\diff v_3}{
%\diffx_1\diff x_2\diff x_3
\int\int\int f(x_1,x_2,x_3,v_1,
v_2,v_3,t)\diff v_1\diff v_2\diff v_3}~~;
\end{lefteqnarray}
or, using Eq.\,(\ref{eq:n}):
\begin{lefteqnarray}
\label{eq:vpn}
&& \overline{v_p}(x_1,x_2,x_3,t)=\frac{\int\int\int
f(x_1,x_2,x_3,v_1,v_2,v_3,t)v_p\diff v_1\diff v_2
\diff v_3}{n(x_1,x_2,x_3,t)}~~; \\
\label{eq:vpqn}
&& \overline{v_pv_q}(x_1,x_2,x_3,t)=\frac{\int\int
\int f(x_1,x_2,x_3,v_1,v_2,v_3,t)v_pv_q\diff v_1\diff 
v_2\diff v_3}{n(x_1,x_2,x_3,t)}~~;
\end{lefteqnarray}
in terms of the number density, $n$.

Let us define the distribution function in the
velocity space:
\begin{equation}
\label{eq:F}
F(x_1,x_2,x_3,v_1,v_2,v_3,t)=\frac{f(x_1,x_2,x_3,
v_1,v_2,v_3,t)}{n(x_1,x_2,x_3,t)}~~;
\end{equation}
which, owing to Eq.\,(\ref{eq:n}), satisfies
the normalization condition:
\begin{equation}
\label{eq:F1}
\int\int\int F(x_1,x_2,x_3,v_1,v_2,v_3,t)\diff v_1
\diff v_2\diff v_3=1~~;
\end{equation}
and the substitution of Eq.\,(\ref{eq:F}) into
(\ref{eq:vpn}) and (\ref{eq:vpqn}) yields:
\begin{lefteqnarray}
\label{eq:vpF}
&& \overline{v_p}(x_1,x_2,x_3,t)=\int\int\int
F(x_1,x_2,x_3,v_1,v_2,v_3,t)v_p\diff v_1\diff v_2
\diff v_3~~; \\
\label{eq:vpqF}
&& \overline{v_pv_q}(x_1,x_2,x_3,t)=\int\int\int
F(x_1,x_2,x_3,v_1,v_2,v_3,t)v_pv_q\diff v_1\diff 
v_2\diff v_3~~;
\end{lefteqnarray}
in terms of the distribution function, $F$.

From a statistical standpoint, the distribution
function, $F$, may be interpreted as a probability
density in the velocity space, where $F(x_1,x_2,x_3,
v_1,v_2,$ $v_3,t)\diff v_1\diff v_2\diff v_3$ represents
the probability of finding a particle inside the
volume element, $\diff^3S$, at the time, $t$, with
velocity components in the range, $v_r\mp\diff v_r/
2$, $r=1,2,3$.   In this view, the velocity components,
$v_p$, and the product velocity components, $v_pv_q$,
may be considered as random variables.   According to
the general definition of variance and covariance
(e.g., Oliva \& Terrasi 1976, Chap.\,II, \S\,2.7),
the following relations hold:
\begin{lefteqnarray}
\label{eq:vvp}
&& (v_p^2)^\ast=(v_p^\ast)^2+\sigma_{v_p}^2~~; \\
\label{eq:cvpq}
&& (v_pv_q)^\ast=v_p^\ast v_q^\ast+\sigma_{v_p
v_q}~~;
\end{lefteqnarray}
where the asterisk denotes the expectation
value of the related distribution, $\sigma_a^
2$ is the mathematical (intended as opposite
to empirical) variance and $\sigma_{ab}$ is the
mathematical covariance.   According if two
random variables, $a$ and $b$, are independent,
correlated, or anticorrelated, the mathematical
covariance, $\sigma_{ab}$,
is null, positive, or negative, respectively.

With regard to an infinitesimal volume element,
$\diff^3S$, at the time, $t$, the first term on 
the right-hand side of Eqs.\,(\ref{eq:vvp}) and 
(\ref{eq:cvpq}) is related to the velocity
components of the centre of mass, while the 
second term is related to velocity components
with respect to the centre of mass.

Expectation values and mathematical variances
and covariances are a priori quantities which
cannot be determined by data collections.
The related observables are arithmetic means
and empirical variances and covariances (e.g.,
Oliva \& Terrasi 1976, Chap.\,IV, \S\,4.3),
and Eqs.\,(\ref{eq:vvp}) and (\ref{eq:cvpq})
translate into:
\begin{lefteqnarray}
\label{eq:vvm}
&& \overline{(v_p^2)}=(\overline{v_p})^2+
\sigma_{v_p}^2~~; \\
\label{eq:cvm}
&& \overline{v_pv_q}=\overline{v_p}~\overline
{v_q}+\sigma_{v_pv_q}~~;
\end{lefteqnarray}
where the notation of variances and
covariances has been left unchanged,
for sake of simplicity.   Finally,
Eq.\,(\ref{eq:vvm}) may also be conglobed
into Eq.\,(\ref{eq:cvm}), using the
definitions:
\begin{equation}
\label{eq:st}
\overline{v_rv_r}=\overline{(v_r^2)}~~;\qquad
\sigma_{v_rv_r}=\sigma_{v_r}^2~~;
\end{equation}
where $\sigma_{v_pv_q}$ may be considered
as the generic component of an empirical
covariance tensor.   If velocity components
are independent, $\sigma_{v_pv_q}=\delta_
{pq}\sigma_{v_p}^2$, where $\delta_{pq}$
is the Kronecker symbol, and Eqs.\,(\ref
{eq:vvm}) and (\ref{eq:cvm}) merge into:
\begin{equation}
\label{eq:virdm}
\overline{v_pv_q}=\overline{v_p}~\overline{v_q}
+\delta_{pq}\sigma_{pp}^2~~;
\end{equation}
where $\sigma_{rr}^2=\sigma_{v_rv_r}=\sigma_
{v_r}^2$ for simplifying the notation and
considering $\sigma_{rr}$ as velocity
dispersions related to random motions,
with regard to a generic infinitesimal
volume element, $\diff^3S$, at the time,
$t$.

Having the centre of mass of the system
been chosen as origin of the (inertial)
reference frame, mean (over the whole
volume) velocity components along
coordinate axes are necessarily null:
$\overline{v_r}=0;$ (e.g., LL66, Chap.\,II,
\S\,8).   Accordingly, $\sigma_{rr}^2=
\overline{(v_r^2)}$, $\sigma_{pq}^2=0$,
$p\ne q$,
and the random square velocity tensor,
$\sigma_{pq}^2=\overline{v_pv_q}$, is
diagonal.

\subsection{Radial and tangential velocity
dispersion on the equatorial plane}
\label{etrvd}

Let us define the axes of the system as
the intersection between the related
volume and principal axes of inertia,
and semiaxes the distances between the
top axes and the centre of mass.   In
general, semiaxes on opposite sides
%with respect to the barycentre,
are different in length.   Let the
coordinate axis, $x_3$, containing
the axis, $a_3$, be chosen as rotation
axis, without loss of generality, as
the reference frame may arbitrarily
be oriented, provided the origin
coincides with the centre of mass.
Let us define the principal plane,
$({\sf O}x_1x_2)$, as the equatorial
plane of the system.

\begin{figure}
\centering
%\resizebox{\hsize}{!}{\includegraphics{syrf2.eps}} 
\centerline{\psfig{file=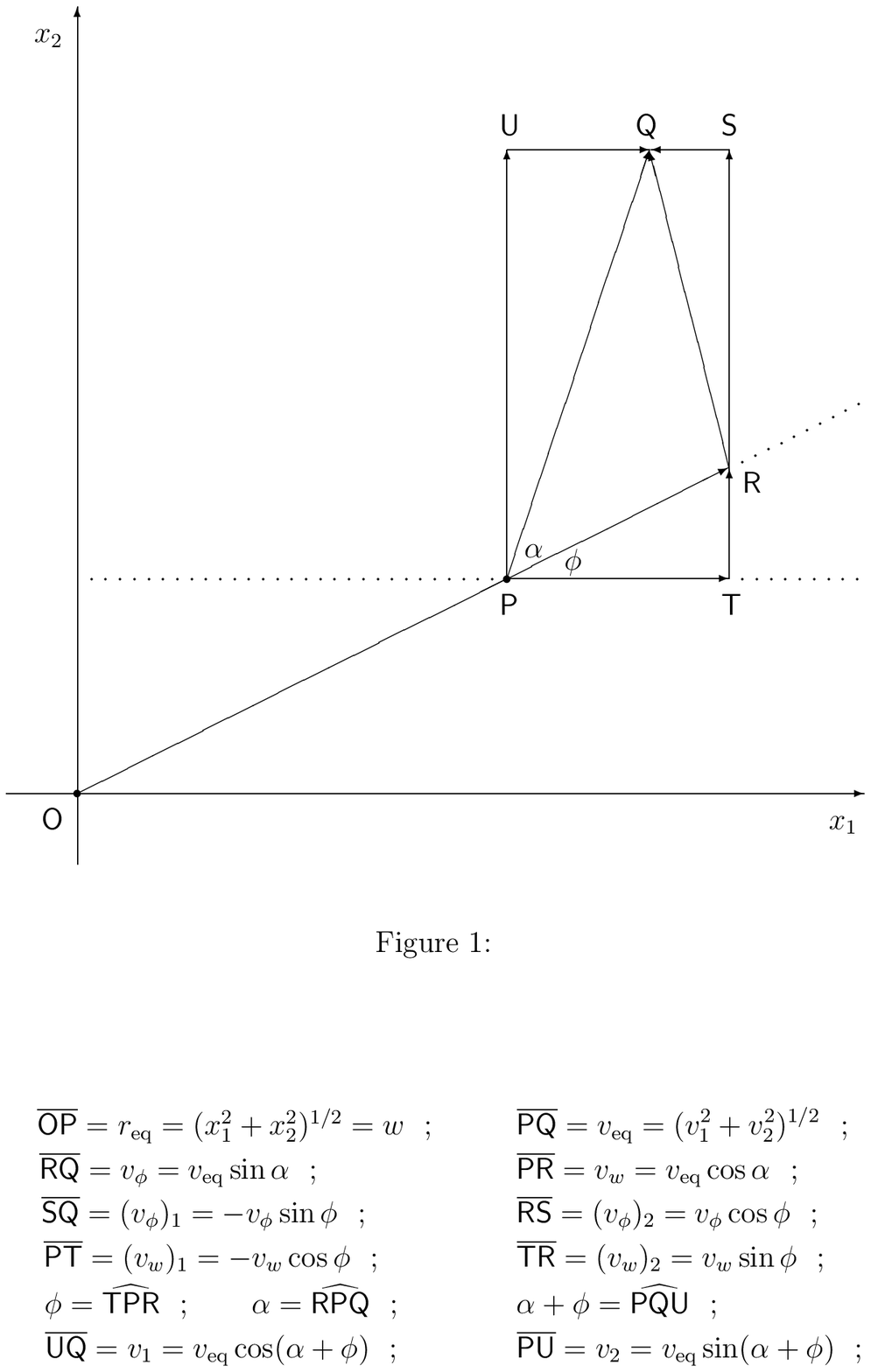,height=180mm,width=150mm}}
\caption{Radial $(v_w)$ and tangential
$(v_\phi)$ velocity components on the
equatorial plane, $({\sf O}x_1x_2)$.
The projection onto the equatorial plane
of the vector radius, $\vec{r}$, and the
velocity, $\vec{v}$, is denoted as $r_{\rm eq}$
and $v_{\rm eq}$, respectively, with regard to
a generic particle.}
\label{f:vphi}
\end{figure}

With regard to a particle placed at a
position, $\vec{r}=(x_1,x_2,x_3)$, and
moving at a velocity,
$\vec{v}=(v_1,v_2,v_3)$, let $\vec{r}_
{\rm eq}=(x_1,x_2)$ and $\vec{v}_{\rm eq}=(v_1,
v_2,)$ be the related projections onto the
equatorial plane, see Fig.\,\ref{f:vphi}.

Cartesian velocity components may be
expressed as the algebraic sum of
radial and tangential velocity
projection on the related direction, as:
\begin{leftsubeqnarray}
\slabel{eq:vcpa}
&& v_1=v_{\rm eq}\cos(\alpha+\phi)=(v_w)_1+
(v_\phi)_1=v_w\cos\phi-v_\phi\sin\phi~~; \\
\slabel{eq:vcpb}
&& v_2=v_{\rm eq}\sin(\alpha+\phi)=(v_w)_2+
(v_\phi)_2=v_w\sin\phi+v_\phi\cos\phi~~;
\label{seq:vcp}
\end{leftsubeqnarray}
conversely, radial and tangential
velocity components may be expressed
as the algebraic sum of cartesian
velocity projections on the related
direction, as:
\begin{leftsubeqnarray}
\slabel{eq:vpca}
&& v_w=v_{\rm eq}\cos\alpha=(v_1)_w+
(v_2)_w=v_1\cos\phi+v_2\sin\phi~~; \\
\slabel{eq:vpcb}
&& v_\phi=v_{\rm eq}\sin\alpha=(v_1)_\phi+
(v_2)_\phi=-v_1\sin\phi+v_2\cos\phi~~;
\label{seq:vpc}
\end{leftsubeqnarray}
where $(v_\mu)_r=[\vec{v}\cdot\vers
(\mu)]\vers(\mu)\cdot\vers(x_r)$;
$(v_r)_\mu=[\vec{v}\cdot\vers(x_r)]
\vers(x_r)\cdot\vers(\mu)$; $\mu=w,
\phi$; $r=1,2$; and $\vers(d)$ is
the unit vector with positive
orientation, along the $d$ direction.

With regard to a generic infinitesimal
volume element, $\diff^3S$, at the time,
$t$, radial and tangential velocity
components, defined by Eqs.\,(\ref
{seq:vpc}), may be considered as
random variables.   Owing to a theorem
of statistics%
\footnote{Let $m_1$, $m_2$, ..., $m_n$,
be random variables and $f_1(m_1)\diff
m_1$, $f_2(m_2)\diff m_2$, ..., $f_n(m_
n)\diff m_n$, related
distributions, $m=\sum_{k=1}^n\alpha_k
m_k$ an additional random variable,
where $\alpha_k$ are coefficients, and
$f(m)\diff m$ a related distribution.
Then the expectation value, $m^\ast$,
is expressible via the above linear
combination of the expectation values,
$m_1^\ast$, $m_2^\ast$, ..., $m_n^\ast$,
as: $m^\ast=\sum_{k=1}^n\alpha_km_k^\ast$.},
the expectation values of the related
distributions read:
\begin{leftsubeqnarray}
\slabel{eq:vpa}
&& v_w^\ast=v_1^\ast\cos\phi+
v_2^\ast\sin\phi~~; \\
\slabel{eq:vpb}
&& v_\phi^\ast=-v_1^\ast\sin\phi+
v_2^\ast\cos\phi~~;
\label{seq:vp}
\end{leftsubeqnarray}
similarly, the expectation values of
the distributions depending on the
random variables, $v_w^2$ and $v_\phi
^2$, are found to be:
\begin{leftsubeqnarray}
\slabel{eq:v2pa}
&& (v_w^2)^\ast=(v_1^2)^\ast\cos^2\phi+
(v_2^2)^\ast\sin^2\phi+2(v_1v_2)^\ast\cos
\phi\sin\phi~~; \\
\slabel{eq:v2pb}
&& (v_\phi^2)^\ast=(v_1^2)^\ast\sin^2\phi+
(v_2^2)^\ast\cos^2\phi-2(v_1v_2)^\ast\sin
\phi\cos\phi~~;
\label{seq:v2p}
\end{leftsubeqnarray}
and using the general definitions expressed
by Eqs.\,(\ref{eq:vvp}) and (\ref{eq:cvpq}),
the related mathematical variances read:
\begin{leftsubeqnarray}
\slabel{eq:vavpa}
&& \sigma_{v_w}^2=(v_w^2)^\ast-(v_w^\ast)^2=
\sigma_{v_1}^2\cos^2\phi+\sigma_{v_2}^2
\sin^2\phi+2\sigma_{v_1v_2}\cos\phi\sin\phi~~; \\
\slabel{eq:vavpb}
&& \sigma_{v_\phi}^2=(v_\phi^2)^\ast-(v_\phi^
\ast)^2=\sigma_{v_1}^2\sin^2\phi+\sigma_{v_2}^2
\cos^2\phi-2\sigma_{v_1v_2}\sin\phi\cos\phi~~;
\label{seq:vavp}
\end{leftsubeqnarray}
where the mathematical covariance, $\sigma_{v_1v_2}$, is
null provided the related velocity components,
$v_1$ and $v_2$, are independent.   The
validity of the relations:
\begin{lefteqnarray}
\label{eq:eqva}
&& (v_w^\ast)^2+(v_\phi^\ast)^2=
(v_1^\ast)^2+(v_2^\ast)^2~~; \\
\label{eq:eqv2a}
&& (v_w^2)^\ast+(v_\phi^2)^\ast=
(v_1^2)^\ast+(v_2^2)^\ast~~; \\
\label{eq:eqvam}
&& \sigma_{v_w}^2+\sigma_{v_\phi}^2=
\sigma_{v_1}^2+\sigma_{v_2}^2~~;
\end{lefteqnarray}
can easily be checked.

In terms of the related observables,
arithmetic means and empirical variances
and covariances, Eqs.\,(\ref{seq:vp}),
(\ref{seq:v2p}), and (\ref{seq:vavp})
translate into:
\begin{leftsubeqnarray}
\slabel{eq:mpa}
&& \overline{v_w}=\overline{v_1}\cos\phi+
\overline{v_2}\sin\phi~~; \\
\slabel{eq:mpb}
&& \overline{v_\phi}=-\overline{v_1}\sin\phi+
\overline{v_2}\cos\phi~~;
\label{seq:mp}
\end{leftsubeqnarray}
\begin{leftsubeqnarray}
\slabel{eq:m2pa}
&& \overline{(v_w^2)}=\overline{(v_1^2)}\cos^2\phi+
\overline{(v_2^2)}\sin^2\phi+2\overline{v_1v_2}\cos
\phi\sin\phi~~; \\
\slabel{eq:m2pb}
&& \overline{(v_\phi^2)}=\overline{(v_1^2)}\sin^2\phi+
\overline{(v_2^2)}\cos^2\phi-2\overline{v_1v_2}\sin
\phi\cos\phi~~;
\label{seq:m2p}
\end{leftsubeqnarray}
\begin{leftsubeqnarray}
\slabel{eq:mawpa}
&& \sigma_{v_w}^2=\overline{(v_w^2)}-(\overline
{v_w})^2=\sigma_{v_1}^2\cos^2\phi+\sigma_{v_2}^2
\sin^2\phi+2\sigma_{v_1v_2}\cos\phi\sin\phi~~; \\
\slabel{eq:mawpb}
&& \sigma_{v_\phi}^2=\overline{(v_\phi^2)}-
(\overline{v_\phi})^2=\sigma_{v_1}^2\sin^2\phi+
\sigma_{v_2}^2\cos^2\phi-2\sigma_{v_1v_2}\sin\phi
\cos\phi~~;
\label{seq:mawp}
\end{leftsubeqnarray}
where the notation of variances and
covariances has been left unchanged,
for sake of simplicity.  The validity
of the relations:
\begin{lefteqnarray}
\label{eq:eqvm}
&& (\overline{v_w})^2+(\overline{v_\phi})^2=
(\overline{v_1})^2+(\overline{v_2})^2~~; \\
\label{eq:eqv2m}
&& \overline{(v_w^2)}+\overline{(v_\phi^2)}=
\overline{(v_1^2)}+\overline{(v_2^2)}~~; \\
\label{eq:eqvar}
&& \sigma_{v_w}^2+\sigma_{v_\phi}^2=
\sigma_{v_1}^2+\sigma_{v_2}^2~~;
\end{lefteqnarray}
can easily be checked.

If velocity components are independent,
$\sigma_{v_pv_q}=\delta_{pq}\sigma_{v_p}^2$,
Eqs.\,(\ref{seq:mawp}) reduce to:
\begin{leftsubeqnarray}
\slabel{eq:vawpa}
&& \sigma_{ww}^2=\sigma_{11}^2\cos^2\phi+
\sigma_{22}^2\sin^2\phi~~; \\
\slabel{eq:vawpb}
&& \sigma_{\phi\phi}^2=\sigma_{11}^2\sin^2\phi+
\sigma_{22}^2\cos^2\phi~~;
\label{seq:vawp}
\end{leftsubeqnarray}
where $\sigma_{rr}^2=\sigma_{v_rv_r}=\sigma_
{v_r}^2$ for simplifying the notation and
considering $\sigma_{rr}$ as velocity
dispersions related to random motions with
regard to a generic infinitesimal volume
element, $\diff^3S$, at the time, $t$.

With regard to the whole volume, $S$, at
the time, $t$, let us define positive
and negative equatorial radial velocity
components, $v_w$, as directed outwards
and inwards, respectively, and positive
and negative equatorial tangential
velocity components, $v_\phi$, as
related to counterclockwise and
clockwise motion, respectively,
around the rotation axis.

Owing to the above mentioned
theorem of statistics, the following relations
hold for for expectation values and
mathematical variances
related to the distributions depending on
radial and tangential velocity
components on the equatorial plane:
\begin{leftsubeqnarray}
\slabel{eq:vpSa}
&& v_w^\ast=\frac1M\int\int\int[v_w(x_1,x_2,x_3,t)]^
\ast\rho(x_1,x_2,x_3,t)\diff x_1\diff x_2\diff x_3~~; \\
\slabel{eq:vpSb}
&& v_\phi^\ast=\frac1M\int\int\int[v_\phi(x_1,x_2,x_3,t)]^
\ast\rho(x_1,x_2,x_3,t)\diff x_1\diff x_2\diff x_3~~;
\label{seq:vpS}
\end{leftsubeqnarray}
\begin{leftsubeqnarray}
\slabel{eq:v2Sa}
&& (v_w^2)^\ast=\frac1M\int\int\int[v_w^2(x_1,x_2,x_3,t)]^
\ast\rho(x_1,x_2,x_3,t)\diff x_1\diff x_2\diff x_3~~; \\
\slabel{eq:v2Sb}
&& (v_\phi^2)^\ast=\frac1M\int\int\int[v_\phi^2(x_1,x_2,x_3,t)]^
\ast\rho(x_1,x_2,x_3,t)\diff x_1\diff x_2\diff x_3~~;
\label{seq:v2S}
\end{leftsubeqnarray}
\begin{leftsubeqnarray}
\slabel{eq:varSa}
&& \sigma_{v_w}^2=(v_w^2)^\ast-(v_w^\ast)^2~~; \\
\slabel{eq:varSb}
&& \sigma_{v_\phi}^2=(v_\phi^2)^\ast-(v_\phi^\ast)^2~~;
\label{seq:varS}
\end{leftsubeqnarray}
where the validity of  Eqs.\,(\ref{eq:eqva}),
(\ref{eq:eqv2a}), and (\ref{eq:eqvam}) is left
unchanged.

In terms of the related observables, arithmetic
means and empirical variances, Eqs.\,(\ref
{seq:vpS}), (\ref{seq:v2S}), and (\ref{seq:varS}),
translate into:
\begin{leftsubeqnarray}
\slabel{eq:mpSa}
&& \overline{v_w}=\frac1 M\int\int\int\overline{v_w}(x_1,x_2,
x_3,t)\rho(x_1,x_2,x_3,t)\diff x_1\diff x_2\diff x_3~~; \\
\slabel{eq:mpSb}
&& \overline{v_\phi}=\frac1 M\int\int\int\overline{v_\phi}
(x_1,x_2,x_3,t)\rho(x_1,x_2,x_3,t)\diff x_1\diff x_2\diff x_3~~;
\label{seq:mpS}
\end{leftsubeqnarray}
\begin{leftsubeqnarray}
\slabel{eq:m2Sa}
&& \overline{v_w^2}=\frac1 M\int\int\int\overline{v_w^2}
(x_1,x_2,x_3,t)\rho(x_1,x_2,x_3,t)\diff x_1\diff x_2\diff x_3~~; \\
\slabel{eq:m2Sb}
&& \overline{v_\phi^2}=\frac1M\int\int\int\overline{v_\phi^2}
(x_1,x_2,x_3,t)]\rho(x_1,x_2,x_3,t)\diff x_1\diff x_2\diff x_3~~;
\label{seq:m2S}
\end{leftsubeqnarray}
\begin{leftsubeqnarray}
\slabel{eq:marSa}
&& \sigma_{v_w}^2=\sigma_{ww}^2=\overline{(v_w^2)}-(\overline
{v_w})^2~~; \\
\slabel{eq:marSb}
&& \sigma_{v_\phi}^2=\sigma_{\phi\phi}^2=\overline{(v_\phi^2)}-
\overline{(v_\phi)}^2~~;
\label{seq:marS}
\end{leftsubeqnarray}
where the validity of Eqs.\,(\ref{eq:eqvm}),
(\ref{eq:eqv2m}), and (\ref{eq:eqvar}) is left
unchanged.

With regard to the angular velocity, $\Omega$,
and the related moment of inertia, $I_3$, the
counterparts
of Eqs.\,(\ref{eq:mpSb}) and (\ref{eq:m2Sb}) read:
\begin{lefteqnarray}
\label{eq:O}
&& \overline{\Omega}=\frac1{I_3}
\int\int\int\overline{\Omega}(x_1,x_2,x_3,t)
w^2\rho(x_1,x_2,x_3,t)\diff x_1\diff x_2\diff
x_3~~; \\
\label{eq:O2}
&& \overline{\Omega^2}=\frac1{I_3}
\int\int\int\overline{\Omega^2}(x_1,x_2,x_3,t)
w^2\rho(x_1,x_2,x_3,t)\diff x_1\diff x_2\diff
x_3~~; \\
\label{eq:Ove}
&& \sigma_\Omega^2=\overline{\Omega^2}-
\overline{\Omega}^2~~; \\
%\label{eq:LO}
%&& L=\int\int\int w\rho(x_1,x_2,x_3,t)\diff
%x_1\diff x_2\diff x_3~~; \\
\label{eq:IO}
&& I_3=\int\int\int w^2\rho(x_1,x_2,x_3,t)\diff
x_1\diff x_2\diff x_3~~;
\end{lefteqnarray}
where the
angular velocity, $\overline{\Omega}$, may be
conceived as a figure rotation, in the sense
that the mean is null when performed in a
reference frame in rigid rotation at the same
rate.   To this respect, particles with
different masses must uniformly be distributed
within the region of phase hyperspace accessible
to the system.

Owing to Eqs.\,(\ref{eq:varSb}), (\ref{eq:O2}),
and (\ref{eq:Ove}), the following relation
holds:
\begin{lefteqnarray}
\label{eq:Od2}
&& M(\overline{v_\phi}^2+\sigma_{\phi\phi}^2)=M
\overline{v_\phi^2}=I_3\overline{\Omega^2}=I_3
(\overline{\Omega}^2+\sigma_\Omega^2)~~;
\end{lefteqnarray}
where the mean angular velocity, $\overline
{\Omega}$, and the empirical variance, $\sigma_
\Omega$, are related to systematic and random 
motions, respectively, around the rotation axis.
Accordingly, Eq.\,(\ref{eq:Od2}) may be splitted
as:
\begin{leftsubeqnarray}
\slabel{eq:vpOa}
&& M\overline{v_\phi}^2=I_3\overline{\Omega}^2~~; \\
\slabel{eq:vpOb}
&& M\sigma_{\phi\phi}^2=I_3\sigma_{\Omega}^2~~;
\label{seq:vpO}
\end{leftsubeqnarray}
which express the contribution of systematic and
random motions along the equatorial plane, in
terms of tangential and angular velocities.

The mean radial velocity component, $\overline{v_w}$,
is related to the motion of the centre of mass along
the equatorial plane.   On the other hand, the centre
of mass coincides with the origin of the coordinates,
which implies $\overline{v_w}=0$ and, in turn,
$\overline{(v_w^2)}=\sigma_{ww}^2$.

The mean tangential velocity component, $\overline
{v_\phi}$, and the empirical variance, $\sigma_
{\phi\phi}^2$, are related to systematic and random motions,
respectively, around the rotation axis.   The diagonal
components of the kinetic-energy tensor may be 
expressed in terms of the above mentioned contributions,
as:
\begin{equation}
\label{eq:Trr}
T_{kk}=(T_{\rm sys})_{kk}+(T_{\rm rdm})_{kk}~~;
\end{equation}
where $k=w,\phi$ in the case under discussion, and:
\begin{leftsubeqnarray}
\slabel{eq:Twra}
&& (T_{\rm sys})_{ww}=0~~; \\
\slabel{eq:Twrb}
&& (T_{\rm rdm})_{ww}=\frac12M\sigma_{ww}^2~~;
\label{seq:Twr}
\end{leftsubeqnarray}
\begin{leftsubeqnarray}
\slabel{eq:Tpra}
&& (T_{\rm sys})_{\phi\phi}=\frac12I_3\overline
{\Omega}^2~~; \\
\slabel{eq:Tprb}
&& (T_{\rm rdm})_{\phi\phi}=\frac12I_3\sigma_
\Omega^2~~;
\label{seq:Tpr}
\end{leftsubeqnarray}
where the indices, sys and rdm, denote
systematic and random motions, respectively.

The global contribution:
\begin{equation}
\label{eq:OT}
T_{\phi\phi}=\frac12I_3\overline{\Omega^2}~~;
\end{equation}
depends only
on the mass distribution, via the moment of
inertia, $I_3$, related to the rotation
axis, $x_3$, and the tangential velocity
component on the equatorial plane, via the
rms angular velocity, $\overline{\Omega^2}$,
regardless from the amount of systematic and
random motions along the direction under
discussion (e.g., Meza 2002; C06).

The contribution of the kinetic-energy
tensor component, $T_{\phi\phi}$, to the
kinetic-energy tensor components, $T_{11}$
and $T_{22}$, owing to Eqs.\,(\ref{seq:Tpr})
and (\ref{eq:OT}), is:
\begin{lefteqnarray}
\label{eq:Tpprr}
&& (T_{\phi\phi})_{qq}=\frac12I_{qq}\overline
{\Omega^2}~~;\qquad q=1,2~~; \\
\label{eq:Tsprr}
&& [(T_{\rm sys})_{\phi\phi}]_{qq}=\frac12I_
{qq}\overline{\Omega}^2~~;\qquad q=1,
2~~; \\
\label{eq:Trprr}
&& [(T_{\rm rdm})_{\phi\phi}]_{qq}=\frac12
I_{qq}(\overline{\Omega^2}-\overline{\Omega}
^2)~~;\qquad q=1,2~~; \\
\label{eq:Lrr}
&& I_{pq}=\int\int\int x_px_q\rho(x_1,x_2,
x_3,t)\diff x_1\diff x_2\diff x_3~~; \\
%\label{eq:L12}
%&& L=I_{11}+I_{22}~~; \\
\label{eq:I12}
&& I_3=I_{11}+I_{22}~~;
\end{lefteqnarray}
where $I_{pq}$ is the moment of inertia tensor.

With regard to equatorial radial kinetic-energy
tensor components, the combination of Eqs.\,(\ref
{eq:eqvar}), (\ref{eq:Ove}), and (\ref{seq:vpO})
yields:
\begin{equation}
\label{eq:vw12p}
\sigma_{ww}^2=\sigma_{11}^2+\sigma_{22}^2-\frac{I_3}M
(\overline{\Omega^2}-\overline{\Omega}^2)~~;
\end{equation}
and the contribution of the kinetic-energy
tensor component, $T_{ww}$, to the kinetic-energy
tensor components, $T_{11}$, $T_{22}$, owing to
Eqs.\,(\ref{eq:Trr}) and (\ref{eq:I12}), is:
\begin{equation}
\label{eq:TwrO}
(T_{ww})_{qq}=\frac12M\sigma_{qq}^2-\frac12
I_{qq}(\overline{\Omega^2}-\overline{\Omega}^2)
~~;\qquad q=1,2~~;
\end{equation}
to be used together with Eq.\,(\ref{eq:Tpprr}).

It is worth noticing that anisotropic random
velocity components distributions, $\sigma_{TT}
\ne\sigma_{RR}$, are not necessarily related
to the shape of the system, while $\sigma_{WW}
\ne2\sigma_{33}$ are (e.g., BT87, Chap.\,4,
\S\,3), where $T$, $R$, and $W$, denote
tangential, radial, and equatorial velocity
components, respectively.   In fact, a
spherically symmetric mass distribution
could, in principle, allow purely radial
or circular orbits.

\subsection{Virial equilibrium configurations}
\label{viec}

The particularization of Eq.\,(\ref{eq:virdI})
to Newtonian interaction, $\chi=-1$, after
combination with Eqs.\,(\ref{eq:Tpprr}) and
(\ref{eq:TwrO}), allows the formulation of the
virial equations for the case under discussion.
The result is:
\begin{leftsubeqnarray}
\slabel{eq:virLa}
&& \frac12\ddot{I}_{qq}=I_{qq}\overline
{\Omega}^2+M\sigma_{qq}^2+(E_{\rm pot})_{qq}~~;
\qquad q=1,2~~; \\
\slabel{eq:virLb}
&& \frac12\ddot{I}_{33}=M\sigma_{33}^2+(E_{\rm pot})
_{33}~~;
\label{seq:virL}
\end{leftsubeqnarray}
where $\overline{v_3}=0$, the system centre
of mass having been chosen as origin of the
reference frame.

The virial equations of the second order,
expressed by Eqs.\,(\ref{eq:virdI}), in
particular Eqs.\,(\ref{seq:virL}), imply
the validity of the following assumptions.
\begin{description}%                          Beccari,  p. 26
\item[\rm{(i)}\hspace{0.2mm}] The mechanical
system under consideration (in particular
a collisionless, self-gravitating fluid)
is isolated (e.g., LL66, Chap.\,I, \S\,5),
which implies angular moment conservation
(e.g., LL66, Chap.\,II, \S\,9).
\item[\rm{(ii)}~~] The potential energy is
a homogeneous function of the coordinates
with degree, $\chi$ (in particular, $\chi
=-1$).
\end{description}
The validity of the further assumption:
\begin{description}
\item[\rm{(iii-a)}] The generic component
of the moment of inertia tensor depends
linearly on time, according to Eq.\,(\ref
{eq:dI20});
\end{description}
or, alternatively:
\begin{description}
\item[\rm{(iii-b)}] The first time
derivative of the generic component
of the moment of inertia tensor is
a bounded function, according to
Eq.\,(\ref{eq:Ilim});
\end{description}
makes Eqs.\,(\ref{seq:virL}) reduce to:
\begin{leftsubeqnarray}
\slabel{eq:vita}
&& I_{qq}\overline{\Omega}^2+M
\sigma_{qq}^2+(E_{\rm pot})_{qq}=0~~;
\qquad q=1,2~~; \\
\slabel{eq:vitb}
&& M\sigma_{33}^2+(E_{\rm pot})_{33}=0~~;
\label{seq:vit}
\end{leftsubeqnarray}
where the variables are to be intended
as instantaneous or averaged over a
sufficiently long time, according
if assumption (iii-a) or (iii-b),
respectively, has been chosen.

A more general formulation of
Eqs.\,(\ref{seq:vit}), which includes
instantaneuos configurations under
assumption (iii-b), is:
\begin{leftsubeqnarray}
\slabel{eq:viza}
&& I_{qq}\overline{\Omega}^2+M
\zeta_{qq}\sigma^2+(E_{\rm pot})_{qq}=0~~;
\qquad q=1,2~~; \\
\slabel{eq:vizb}
&& M\zeta_{33}\sigma^2+(E_{\rm pot})_{33}=0~~; \\
\slabel{eq:vizc}
&& \sigma^2=\sigma_{11}^2+\sigma_{22}^2+
\sigma_{33}^2~~; \\
\slabel{eq:vizd}
&& \zeta_{rr}=\frac{(\tilde{E}_{{\rm rdm}})_
{rr}}{E_{{\rm rdm}}}=\frac{\tilde{\sigma}_{rr}^
2}{\sigma^2}~~;\qquad r=1,2,3~~; \\
\slabel{eq:vize}
&& \zeta_{11}+\zeta_{22}+\zeta_{33}=\frac
{\tilde{E}_{{\rm rdm}}}{E_{{\rm rdm}}}=\frac
{\tilde{\sigma}^2}{\sigma^2}=\zeta~~;
\label{seq:viz}
\end{leftsubeqnarray}
where $\zeta_{rr}$ may be conceived as
anisotropy parameters (CM05, C06), $E_
{\rm rdm}$ is the random kinetic energy,
and $\tilde{E}_{\rm rdm}$ is the effective
random kinetic energy i.e. the right
amount needed for an instantaneous
configuration to satisfy Eqs.\,(\ref
{seq:vit}).   Generalized anisotropy
parameters lower or larger than $\zeta/3$
imply, respectively, lack or excess of
random
%(i.e. other than systematic
%rotation around a fixed axis)
motions along the related direction.
On the other hand, the ratios:
\begin{leftsubeqnarray}
\slabel{eq:zita}
&& \tilde{\zeta}_{rr}=\frac{\tilde{E}_
{{\rm rdm}}}{E_{{\rm rdm}}}=\frac{\zeta
_{rr}}{\zeta}~~;\qquad r=1,2,3~~; \\
\slabel{eq:zitb}
&& \tilde{\zeta}_{11}+\tilde{\zeta}_{22}+
\tilde{\zeta}_{33}=1~~;
\label{seq:zit}
\end{leftsubeqnarray}
may be conceived as effective anisotropy
parameters (CM05, C06).

The parameter, $\zeta$, may be conceived
as a virial index, where $\zeta=1$
corresponds to null virial excess,
$2\Delta E_{{\rm rdm}}=2(\tilde{E}_
{{\rm rdm}}-E_{{\rm rdm}})$, which does
not necessarily imply a relaxed
configuration%
\footnote{For instance, a homogeneous
sphere undergoing coherent oscillations
exhibits $\zeta>1$ at expansion
turnover and $\zeta<1$ at contraction
turnover.   Then it necessarily exists
a configuration where $\zeta=1$ which,
on the other hand, is unrelaxed.},
$\zeta>1$ to positive virial excess,
and $\zeta<1$ to negative virial excess.
The special case, $\zeta=1$, makes
Eqs.\,(\ref{seq:viz}) reduce to (\ref
{seq:vit}).

For sake of simplicity, let us define
ideal, self-gravitating fluids, rotating
around an axis, $a_3$, for which 
Eqs.\,(\ref{seq:viz}) hold i.e.
assumptions (i), (ii), and (iii) above
are valid, as R3 fluids.

The combination of Eqs.\,(\ref{eq:viza})
and (\ref{eq:vizb}) yields:
\begin{equation}
\label{eq:vz12}
I_{qq}=\overline{\Omega}^2-\frac{\zeta_
{qq}}{\zeta_{33}}(E_{\rm pot})_{33}+
(E_{\rm pot})_{qq}=0~~;\qquad q=1,2~~;
\end{equation}
to get further insight, let us express
the self potential-energy tensor,
$(E_{\rm pot})_{rr}$, and the moment
of inertia
tensor, $I_{rr}$, in terms of dimensionless
tensors, ${\cal P}_{rr}$ and ${\cal I}_{rr}$,
respectively, as:
\begin{lefteqnarray}
\label{eq:Prra}
&& (E_{\rm pot})_{rr}=-\frac{GM^2}a{\cal P}_{rr}~~;\qquad
(E_{\rm pot})=-\frac{GM^2}a{\cal P}~~;\qquad r=1,2,3~~; \\
\label{eq:Lrra}
&& I_{qq}=Ma^2{\cal I}_{qq}~~;\qquad I_3=Ma^2{\cal I}_3
~~;\qquad q=1,2~~; \\
\label{eq:aref}
&& a=\left(\frac S{2\pi}\right)^{1/3}~~;
\end{lefteqnarray}
in addition, let us define the rotation
parameter:
\begin{equation}
\label{eq:up}
\upsilon=\frac{a^3\overline{\Omega}^2}{GM}=
\frac{\overline{\Omega}^2}{2\pi G\overline
{\rho}}~~;
\end{equation}
where $\overline{\rho}=M/S$ is the mean
density of the system.   In the special
case of solid-body rotation $(\overline
{\Omega}=\Omega)$, Eq.\,(\ref{eq:up})
reduces to a notation used for polytropes
(e.g., Jeans 1929, Chap.\,IX, \S\,232;
Chandrasekhar \& Leboviz 1962), and in
the limit of ellipsoidal homogeneous
configurations $(\overline{\rho}=\rho)$,
Eqs.\,(\ref{eq:up}) reduces to a notation
used for MacLaurin spheroids and Jacobi
ellipsoids (e.g., Jeans 1929, Chap.\,VIII,
\S\S\,189-193; C69, Chap.\,5, \S\,32,
Chap.\,6, \S\,39).

Owing to Eqs.\,(\ref{eq:Prra}), (\ref{eq:Lrra})
and (\ref{eq:aref}), Eq.\,(\ref{eq:vz12})
may be formulated in terms of dimensionless
parameters, as:
\begin{equation}
\label{eq:vza}
(\zeta_{33}{\cal P}_{qq}-\zeta_{qq}{\cal P}_{33})-
\upsilon\zeta_{33}{\cal I}_{qq}=0~~;\qquad
q=1,2~~;
\end{equation}
which admits real solutions provided the
inequality:
\begin{equation}
\label{eq:cone}
\frac{\zeta_{qq}}{\zeta_{33}}\le\frac{{\cal P}_
{qq}}{{\cal P}_{33}}~~;\qquad q=1,2~~;
\end{equation}
is satisfied, that is the natural extension to
R3 fluids of its counterparts related to axisymmetric,
relaxed mass distributions (Wiegandt1982a,b) and
homeoidally striated ellipsoids (C06).   Imaginary
solutions correspond to imaginary rotation
parameters i.e. imaginary rotation.

The combination of Eqs.\,(\ref{eq:vza})
yields after some algebra:
\begin{equation}
\label{eq:vzL}
{\cal I}_{22}(\zeta_{33}{\cal P}_{11}-\zeta_{11}
{\cal P}_{33})={\cal I}_{11}(\zeta_{33}{\cal P}_
{22}-\zeta_{22}{\cal P}_{33})~~;
\end{equation}
or equivalently:
\begin{equation}
\label{eq:vzP}
\zeta_{33}({\cal I}_{22}{\cal P}_{11}-{\cal I}_{11}
{\cal P}_{22})={\cal P}_{33}({\cal I}_{22}\zeta_{11}-
{\cal I}_{11}\zeta_{22})~~;
\end{equation}
which make alternative expressions of the
constraint related to virial equilibrium.

\subsection{Axisymmetric and triaxial
configurations}\label{axtr}

An explicit expression of the rotation
parameter, $\upsilon$, can be derived
from Eqs.\,(\ref{eq:vza}), as:
\begin{equation}
\label{eq:uppq}
\upsilon=\frac{\zeta_{33}{\cal P}_{qq}-\zeta_{qq}
{\cal P}_{33}}{\zeta_{33}{\cal I}_{qq}}
~~;\qquad q=1,2~~;
\end{equation}
which, in turn, allows an explicit
expression of anisotropy parameter ratios,
$\zeta_{pp}/\zeta_{qq}$, as:
\begin{lefteqnarray}
\label{eq:zq3}
&& \frac{\zeta_{qq}}{\zeta_{33}}=\frac
{{{\cal P}}_{qq}}{{{\cal P}}_{33}}
\left[1-\upsilon\frac{{{\cal I}}_{qq}}
{{{\cal P}}_{qq}}\right]~~;\quad q=1,2~~; \\
\label{eq:z12}
&& \frac{\zeta_{11}}{\zeta_{22}}=\frac
{{{\cal P}}_{11}-\upsilon{\cal I}_{11}}
{{{\cal P}}_{22}-\upsilon{\cal I}_{22}}~~;
\end{lefteqnarray}
and the combination of Eqs.\,(\ref{eq:vize}) 
and (\ref{eq:zq3}) yields:
\begin{equation}
\label{eq:z3}
\frac{\zeta_{33}}{\zeta}=\frac{{{\cal P}}_{33}}
{{{\cal P}}-\upsilon{\cal I}_3}~~;
\end{equation}
which provides an alternative expression to
Eqs.\,(\ref{eq:uppq}), as:
\begin{lefteqnarray}
\label{eq:upz}
&& \upsilon=\frac{\zeta_{33}{\cal P}-\zeta
{\cal P}_{33}}{\zeta_{33}{\cal I}_3}~~;
\end{lefteqnarray}
that is equivalent to Eq.\,(\ref{eq:vza}), and
then admit real solutions provided inequality
(\ref{eq:cone}) is satisfied.

Finally, Eqs.\,(\ref{eq:vza}) may be combined as:
\begin{equation}
\label{eq:R12}
\frac{{\cal I}_{11}}{{\cal I}_{22}}=\frac
{\zeta_{33}{{\cal P}}_{11}-\zeta_{11}
{{\cal P}}_{33}}{\zeta_{33}{{\cal P}}_{22}-
\zeta_{22}{{\cal P}}_{33}}~~;
\end{equation}
where it can be seen that Eqs.\,(\ref
{eq:z12}) and (\ref{eq:R12}) are changed
one into the other, by replacing the
terms, ${{\cal P}}_{33}\zeta_{qq}/\zeta_
{33}$, with the terms, $\upsilon{\cal I}_{qq}$,
and vice versa.   The above results may be
reduced to a single statement.
\begin{trivlist}
\item[\hspace\labelsep{\bf Theorem 1.}] \sl
Given a R3 fluid, the relation:
\begin{lefteqnarray*}
&& \frac{X_{11}}{X_{22}}=\frac{{{\cal P}}_{11}
-Y_{11}}{{{\cal P}}_{22}-Y_{22}}~~; \\
&& X_{qq}=\frac{\zeta_{qq}}{\zeta_{33}}{{\cal
P}}_{33}~,~\upsilon{\cal I}_{qq}~~;\qquad
q=1,2~~; \\
&& Y_{qq}=\upsilon{\cal I}_{qq}~,~\frac
{\zeta_{qq}}{\zeta_{33}}{\cal P}_{33}~~;
\qquad q=1,2~~;
\end{lefteqnarray*}
is symmetric with respect to $X_{qq}$ and
$Y_{qq}$, being the former tensor related
to anisotropic random velocity distribution,
and the latter one to systematic rotation
around the axis, $x_3$, or vice versa.
\end{trivlist}

In the special case of axisymmetric configurations,
the dimensionless factors appearing in the
expression of the self potential-energy tensor,
Eqs.\,(\ref{eq:Prra}), and the moment of
inertia, Eqs.\,(\ref{eq:Lrra}), do coincide
with regard to equatorial axes, ${{\cal P}}_{11}=
{{\cal P}}_{22}$ and ${\cal I}_{11}={\cal I}_
{22}$, respectively, which necessarily imply
$\zeta_{11}=\zeta_{22}$, owing to Eq.\,(\ref
{eq:R12}).

In the general case of triaxial configurations,
the contrary holds, ${{\cal P}}_{11}\ne{{\cal P}}_
{22}$ and ${\cal I}_{11}\ne{\cal I}_{22}$,
then the equality,
$\zeta_{11}=\zeta_{22}$, via Eq.\,(\ref
{eq:z12}), implies the validity of the relation:
\begin{equation}
\label{eq:up12}
\upsilon=\frac{{\cal P}_{11}-{\cal P}_{22}}
{{\cal I}_{11}-{\cal I}_{22}}~~;
\end{equation}
if otherwise, the random velocity distribution
along the equatorial plane%
\footnote{Throughout this paper, ``along the
equatorial plane'' has to be intended as
``along any direction parallel to the equatorial
plane''.}
is anisotropic i.e.
$\zeta_{11}\ne\zeta_{22}$.   The related degeneracy
can be removed using an additional condition,
as it will be shown in the next section.

The above results may be reduced to a single
statement.
\begin{trivlist}
\item[\hspace\labelsep{\bf Theorem 2.}] \sl
%Homeoidally striated MacLaurin spheroids
%necessarily imply isotropic random velocity
%distributions along the equatorial plane,
%$\zeta_{11}=\zeta_{22}$.  Homeoidally striated
%Jacobi ellipsoid necessarily imply either
%anisotropic random velocity distribution
%along the equatorial plane, $\zeta_{11}\ne
%\zeta_{22}$, or isotropic random velocity
%distribution, $\zeta_{11}=\zeta_{22}=\zeta_
%{33}=\zeta/3$.
Isotropic random velocity distribution
along the equatorial plane, $\zeta_{11}=
\zeta_{22}$, makes a necessary condition
for R3 fluids to be symmetric with respect
to the rotation axis, $x_3$.
\end{trivlist}

\section{Imaginary rotation}
\label{imro}

A unified theory of systematic and random
motions is allowed, taking into consideration
imaginary rotation.   It has been shown above
that Eq.\,(\ref{eq:vza}), or equivalently
one among (\ref{eq:uppq}), (\ref{eq:upz}),
admits real solutions provided inequality
(\ref{eq:cone}) is satisfied.   If otherwise,
the rotation parameter,  $\upsilon$,
has necessarily to be negative, which implies,
via Eq.\,(\ref{eq:up}), an
{\it imaginary} figure rotation, $\img\Omega$,
where $\img$ is the imaginary unit.
Accordingly, the related centrifugal potential
takes the general expression:
\begin{equation}
\label{eq:Tpm}
{\cal T}(x_1,x_2,x_3,t)=\frac12\Sgn\left(\frac{{\cal P}_
{qq}}{{\cal P}_{33}}-\frac{\zeta_{qq}}{\zeta_{33}}\right)
[\overline{\Omega}(x_1,x_2,x_3,t)]^2w^2~~;\qquad w^2=x_1^2
+x_2^2~~;
\end{equation}
where $\Sgn$ is the sign function, $\Sgn(\mp\vert
x\vert)=\mp1$, $x\ne0$.
The centrifugal force, $\partial{\cal T}/\partial w$,
is positive or negative according if real or
imaginary rotation occurs, respectively.
Then the net effect
of real rotation is flattening, while the net
effect of imaginary rotation is {\it elongation},
with respect to the rotation axis (Caimmi 1996b; C06).

To get further insight, let us particularize
Eq.\,(\ref{eq:vza}) to the special case of
null rotation $(\upsilon=0)$.   The result is:
\begin{equation}
\label{eq:zq30}
\frac{\zeta_{qq}}{\zeta_{33}}=\frac{{{\cal P}}_{qq}}
{{{\cal P}}_{33}}~~;\qquad \upsilon=0~~;\qquad q=1,2~~;
\end{equation}
where the right-hand side, via Eqs.\,(\ref{eq:Vpqc})
and (\ref{eq:Prra}), depends on the mass distribution only.
Accordingly, the net effect of positive $(\zeta_{qq}/
\zeta_{33}>0)$ or negative $(\zeta_{qq}/\zeta_{33}<0)$
random motion excess along the equatorial
plane is flattening $({\cal P}_{qq}>{\cal P}_{33})$
or elongation  $({\cal P}_{qq}<{\cal P}_{33})$,
respectively.
In what follows, it shall be intended that
random motion excess is related to the
equatorial plane.

\subsection{Random motion excess and rotation}
\label{rero}

In the limit of isotropic random velocity
distribution, $\zeta_{11}=\zeta_{22}=\zeta_
{33}=\zeta/3$, Eqs.\,(\ref{eq:uppq}) and
(\ref{eq:upz}) reduce to:
\begin{lefteqnarray}
\label{eq:uq3i}
&& \upsilon_{{\rm iso}}=\frac{{\cal P}_{qq}-
{\cal P}_{33}}{{\cal I}_{qq}}~~;\qquad q=1,2~~; \\
\label{eq:uqi}
&& \upsilon_{{\rm iso}}=\frac{{\cal P}-3{\cal P}_{33}}
{{\cal I}_3}~~;
\end{lefteqnarray}
where the index, ${\rm iso}$, means isotropic
random velocity distribution.

Accordingly, Eqs.\,(\ref{eq:uppq}) and (\ref{eq:upz})
may be expressed as:
\begin{lefteqnarray}
\label{eq:uNia}
&& \upsilon=\upsilon_{{\rm iso}}-\upsilon_{{\rm ani}}~~; \\
\label{eq:uq3a}
&& \upsilon_{{\rm ani}}=\left(\frac{\zeta_{qq}}
{\zeta_{33}}-1\right)\frac{{\cal P}_{33}}
{{\cal I}_{qq}}~~;\qquad q=1,2~~; \\
\label{eq:u3a}
&& \upsilon_{{\rm ani}}=\left(\frac\zeta{\zeta_{33}}-3
\right)\frac{{\cal P}_{33}}{{\cal I}_3}~~;
\end{lefteqnarray}
where $\upsilon_{{\rm ani}}\ge0$ for oblate-like
configurations, $\zeta_{qq}/\zeta_{33}\ge1$;
$\upsilon_{{\rm ani}}\le0$ for prolate-like
configurations, $\zeta_{qq}/\zeta_{33}\le1$;
and the index, ${\rm ani}$, means contribution from
random motion excess which, in general, makes
an anisotropic random velocity distribution.
Accordingly, positive
or negative random motion excess is related to
real or imaginary rotation, respectively.
                              
Let us rewrite Eq.\,(\ref{eq:uNia}) as:
\begin{equation}
\label{eq:uNo}
\upsilon_{{\rm iso}}=\upsilon+\upsilon_
{{\rm ani}}~~;
\end{equation}
which, owing to Eq.\,(\ref{eq:up}),
is equivalent to:
\begin{equation}
\label{eq:O2i}
\overline{\Omega}_{{\rm iso}}^2=\overline
{\Omega}^2+\Sgn\left(\frac\zeta{\zeta_{33}}-3
\right)\overline{\Omega}_{{\rm ani}}^2~~;
\end{equation}
where positive and negative $\Sgn$ values
correspond to real and imaginary rotation,
respectively.   Then the effect of random
motion excess on the shape
of the system, is virtually indistinguishible
from the effect of additional figure
rotation.   The above
results may be reduced to a single statement.
\begin{trivlist}
\item[\hspace\labelsep{\bf Theorem 3.}] \sl
Given a R3 fluid, the effect of (positive or
negative) random motion excess is
equivalent to an additional (real or imaginary)
figure rotation, $\Sgn(\zeta/\zeta_{33}-3)
\overline{\Omega}_{{\rm ani}}^2$, with regard
to an adjoint configuration where the random
velocity distribution is isotropic.
\end{trivlist}
Accordingly, a R3 fluid with assigned
systematic rotation
and random velocity distribution, with
regard to the shape, is virtually
indistinguishable from an adjoint
configuration of equal density profile,
isotropic random velocity distribution,
and figure rotation deduced from
Eq.\,(\ref{eq:O2i}).

\subsection{Axisymmetric and triaxial
configurations}\label{atri}

The combination of alternative expressions of
the rotation parameter, $\upsilon_{{\rm iso}}$,
defined by Eqs.\,(\ref{eq:uq3i}), yields:
\begin{equation}
\label{eq:upep}
{\cal I}_{11}{\cal P}_{22}-{\cal I}_{22}{\cal P}_
{11}={\cal P}_{33}({\cal I}_{11}-{\cal I}_{22})~~;
\end{equation}
which, for axisymmetric configurations,
${\cal I}_{11}={\cal I}_{22}$, ${\cal P}_
{11}={\cal P}_{22}$, reduces to an
indeterminate form, $0=0$.

The combination of alternative expressions of
the rotation parameter, $\upsilon_{{\rm ani}}$,
defined by Eqs.\,(\ref{eq:uNia}), yields:
\begin{equation}
\label{eq:ziep}
{\cal I}_{11}\zeta_{22}-{\cal I}_{22}\zeta_{11}=
\zeta_{33}({\cal I}_{11}-{\cal I}_{22})~~;
\end{equation}
which, for isotropic random velocity
distributions, reduces to an indeterminate
form, $0=0$.   In addition, axisymmetric
configurations $({\cal I}_{11}={\cal I}_{22})$
necessarily imply isotropic random
velocity distributions along the equatorial
plane, $\zeta_{11}=\zeta_{22}$.

The combination of Eqs.\,(\ref{eq:vize})
and (\ref{eq:ziep}) yields:
\begin{lefteqnarray}
\label{eq:z12L}
&& \zeta_{qq}=\frac{\zeta{\cal I}_{qq}-\zeta_
{33}(2{\cal I}_{qq}-{\cal I}_{pp})}{{\cal I}_3}
~~;\qquad q=1,2~~;\qquad p=2,1~~;
\end{lefteqnarray}
which, for axisymmetric configurations
$({\cal I}_{11}={\cal I}_{22}={\cal I}_3/2)$
reduces to :
\begin{equation}
\label{eq:z11L}
\zeta_{qq}=\frac{{\cal I}_{qq}}{{\cal I}_3}
(\zeta-\zeta_{33})=\frac{\zeta-\zeta_{33}}
2~~;\qquad q=1,2~~;
\end{equation}
and the special case, $\zeta_{33}=\zeta/3$,
reads $\zeta_{11}=\zeta_{22}=\zeta/3$.
            
The limiting configuration, $\zeta_{qq}=0$,
via Eqs.\,(\ref{eq:upep}), necessarily
implies ${\cal I}_{pp}\le{\cal I}_{qq}$,
owing to $\zeta_{pp}\ge0$, $q=1,2,$ $p=2,
1,$ and Eq.\,(\ref{eq:z12L}) reduces to:
\begin{equation}
\label{eq:zqq0}
\zeta{\cal I}_{qq}-\zeta_{33}(2{\cal I}_{qq}-
{\cal I}_{pp})=0~~;\qquad q=1,2~~;\qquad p=2,1~~;
\end{equation}
which, owing to Eq.\,(\ref{eq:vize}), is
equivalent to:
\begin{leftsubeqnarray}
\slabel{eq:z3q0a}
&& \frac{{\cal I}_{pp}}{{\cal I}_{qq}}=\frac
{\zeta_{33}-\zeta_{pp}}{\zeta_{33}}=\frac
{2\zeta_{33}-\zeta}{\zeta_{33}}~~;  \\
\slabel{eq:z3q0b}
&& \zeta_{qq}=0~~;\qquad q=1,2~~;\qquad p=2,1~~;
\label{seq:z3q0}
\end{leftsubeqnarray}
where ${\cal I}_{qq}/{\cal I}_{pp}\ge1$
implies $\zeta_{33}\ge\zeta_{pp}$ and
$\zeta_{33}\ge\zeta/2$.
The above results may
be reduced to the following statements.
\begin{trivlist}
\item[\hspace\labelsep{\bf Theorem 4.}] \sl
Given a R3 fluid, 
the anisotropy parameters along the equatorial
plane, $\zeta_{qq}$, $q=1,2,$ depend on the
diagonal components of the dimensionless
moment of inertia
tensor, ${\cal I}_{qq}$, $q=1,2,$ and the
related expressions coincide, $\zeta_{11}=
\zeta_{22}$, in the limit of axisymmetric
configurations, ${\cal I}_{11}={\cal I}_{22}$.
\end{trivlist}
\begin{trivlist}
\item[\hspace\labelsep{\bf Theorem 5.}] \sl
Given a R3 fluid, a necessary and sufficient
condition for
isotropic random velocity distribution
is that the anisotropy parameter along the
rotation axis attains the value, $\zeta_{33}
=1/3$.
\end{trivlist}
\begin{trivlist}
\item[\hspace\labelsep{\bf Theorem 6.}] \sl
Given a sequence of R3 fluids, the ending point
occurs when the third diagonal component of the
dimensionless self potential-energy tensor is
null, ${\cal P}_{33}=0$, and/or the generalized
anisotropy parameter related to the major
equatorial axis is null, $\zeta_{11}
=0$, which is equivalent to ${\cal I}_{22}/
{\cal I}_{11}=(2-\zeta/\zeta_{33})^{1/2}$.
The related value of the rotation parameter
is $\upsilon={\cal P}_{qq}/{{\cal I}_
{qq}}$, $q=1,2,$ independent of anisotropy
parameters.   The special case
of dynamical (or hydrostatic) equilibrium,
$\zeta=1$, implies centrifugal support along
the major equatorial axis, provided $\zeta_{11}=0$.
\end{trivlist}

Accordingly, with regard to R3 fluids, the
anisotropy parameters along the equatorial
plane, $\zeta_{11}$ and $\zeta_{22}$, cannot
be arbitrarily assigned, but depend on the
dimensionless moment of inertia tensor diagonal
components, ${\cal I}_{11}$ and ${\cal I}_{22}$,
conform to Eqs.\,(\ref{eq:z12L}).   On
the other hand, the knowledge of the
dimensionless moment of inertia tensor
components, ${\cal I}_{11}$ and ${\cal I}_
{22}$, the dimensionless self potential-energy
tensor components, ${\cal P}_{11}$, ${\cal P}_
{22}$ and ${\cal P}_{33}$, together with the
rotation parameter, $\upsilon$, 
allows the determination of the rotation
parameter, $\upsilon_{{\rm ani}}$, via
Eqs.\,(\ref{eq:uq3i}), (\ref{eq:uqi}),
(\ref{eq:uNia}), and then the ratios,
$\zeta_{qq}/\zeta_{33}$, $\zeta_{33}/
\zeta$, via Eqs.\,(\ref{eq:uq3a}),    
(\ref{eq:u3a}), respectively, or the
anisotropy parameter
along the rotation axis, $\zeta_{33}$,
provided the virial index, $\zeta$,
defined by Eq.\,(\ref{eq:vize}), is
assigned.

In conclusion, with regard to R3 fluids
defined by assigned
dimensionless moment of inertia tensor
components, ${\cal I}_{11}$ and ${\cal I}_
{22}$, dimensionless self potential-energy
tensor components, ${\cal P}_{11}$, ${\cal P}_
{22}$ and ${\cal P}_{33}$, rotation parameter,
$\upsilon$, and virial index, $\zeta$,
the anisotropy parameters, $\zeta_{11}$,
$\zeta_{22}$, $\zeta_{33}$, cannot
arbitrarily be fixed, but must be
determined as shown above.

\subsection{Sequences of virial equilibrium
configurations}
\label{virco}

With regard to R3 fluids, it has been shown
above that adjoints configurations are
characterized by (i) centrifugal potential,
${{\cal T}}_{{\rm iso}}(x_1,x_2,x_3)={{\cal
T}}(x_1,x_2,x_3)+{{\cal T}}_{{\rm ani}}(x_1,
x_2,x_3)$, or $\overline{\Omega}^2_{{\rm
iso}}(x_1,x_2,x_3)=\overline{\Omega}^2(x_1,
x_2,x_3)+\Sgn(\zeta/\zeta_{33}-3)\overline
{\Omega}^2_{{\rm ani}}(x_1,x_2,x_3)$,
Eq.\,(\ref{eq:O2i}); and (ii)
isotropic random velocity distribution.
Owing to Theorem 3, a sequence of R3 fluids
coincides with the sequence of adjoints
configurations.   Given a R3 fluid with fixed
components of the dimensionless self
potential-energy tensor, ${{\cal P}}_{11}$,
${{\cal P}}_{22}$, ${{\cal P}}_{33}$,
dimensionless moment of inertia tensor,
${\cal I}_{11}$, ${\cal I}_{22}$,
rotation parameter, $\upsilon$,
and virial index, $\zeta$, the
anisotropy parameters, $\zeta_{11}$,
$\zeta_{22}$, $\zeta_{33}$, are
determined via
Eqs.\,(\ref{eq:uq3i})-(\ref{eq:u3a}) and
(\ref{eq:z12L}).   Negative values of
the rotation parameter, $\upsilon_
{{\rm iso}}$, extend the sequence of
axisymmetric configurations to imaginary
rotation i.e. prolate configurations
where the major axis coincides with the
rotation axis.

Once more owing to Theorem 3, the
bifurcation point of a sequence of
R3 fluids coincides with its counterpart
along the sequence of adjoint
configurations.   Aiming to find a
necessary condition for the occurrence
of a bifurcation point, let us equalize
the alternative expressions of
Eqs.\,(\ref{eq:uppq}).  The result is:
\begin{equation}
\label{eq:rzPL}
\frac{{\cal P}_{11}}{{\cal I}_{11}}-\frac
{\zeta_{11}}{\zeta_{33}}\frac{{\cal P}_{33}}
{{\cal I}_{11}}=
\frac{{\cal P}_{22}}{{\cal I}_{22}}-\frac
{\zeta_{22}}{\zeta_{33}}\frac{{\cal P}_{33}}
{{\cal I}_{22}}~~;
\end{equation}
where the anisotropy parameter ratios,
$\zeta_{qq}/\zeta_{33}$, $q=1,2,$ may
be deduced from Eqs.\,(\ref{eq:z12L}).
After some algebra, Eq.\,(\ref{eq:rzPL})
reads:
\begin{equation}
\label{eq:bife}
\frac{{\cal I}_{11}{\cal P}_{22}-{\cal I}_{22}
{\cal P}_{11}}{{\cal I}_{11}-{\cal I}_{22}}=
{\cal P}_{33}~~;
\end{equation}
and the occurrence of a bifurcation
point has necessarily to satisfy the
relation:
\begin{equation}
\label{eq:bifc}
\lim_{\epsilon_{21}\to1}
\frac{{\cal I}_{11}{\cal P}_{22}-{\cal I}_{22}
{\cal P}_{11}}{{\cal I}_{11}-{\cal I}_{22}}=
{\cal P}_{33}~~;
\end{equation}
where $\epsilon_{21}=a_2/a_1$ is
the ratio of two generic equatorial (perpendicular)
radii, and $\epsilon_{21}\to1$
implies ${\cal I}_{22}\to{\cal I}_
{11}$, ${\cal P}_{22}\to{\cal P}_
{11}$.  Then Eq.\,(\ref{eq:bifc})
is a necessary condition for the
existence of a bifurcation point,
as it selects the sole axisymmetric
configuration which satisfies
Eq.\,(\ref{eq:bife}), regardless
from the values of the anisotropy
parameters, $\zeta_{rr}$, $r=1,2,
3.$   The above results
may be reduced to a single statement.
\begin{trivlist}
\item[\hspace\labelsep{\bf Theorem 7.}] \sl
Given a sequence of R3 fluids, a necessary
condition for the existence of a bifurcation
point from axisymmetric to triaxial
configurations, is independent of the
anisotropy parameters, $\zeta_{rr}$,
$r=1,2,3,$ and coincides with its counterpart
related to the sequence of adjoint
configurations with isotropic random
velocity distribution.
\end{trivlist}
                
\begin{figure}
\centering
%\resizebox{\hsize}{!}{\includegraphics{syso.eps}} 
\centerline{\psfig{file=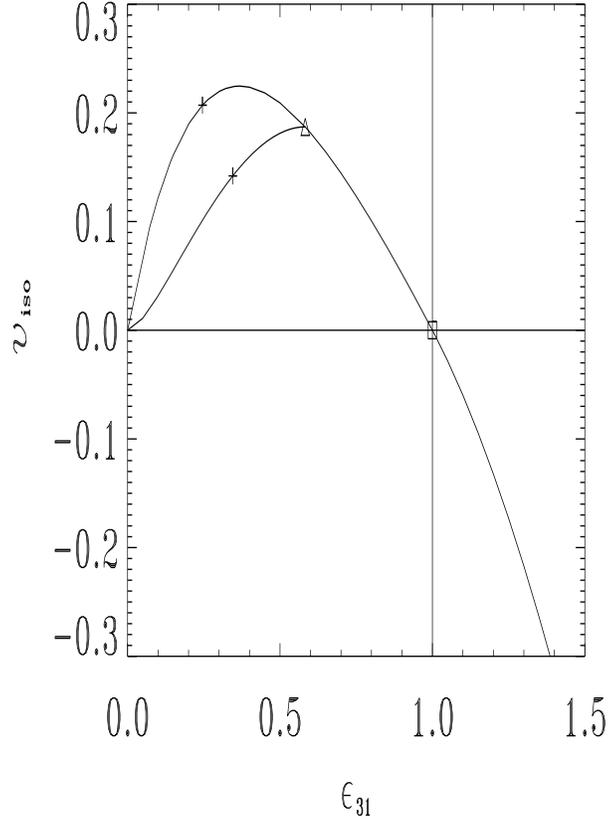,height=130mm,width=90mm}}
\caption{A hypothetical sequence of axisymmetric
R3 fluids, from the starting point (square)
to the bifurcation point (triangle), and
related triaxial R3 fluids, from the
starting point (triangle) to the bifurcation
point (Greek cross), extended to the
case of imaginary rotation (negative values 
of the rotation parameter, $\upsilon$).
In any case, the random velocity distribution
is isotropic.   Both sequences
are continued in the region of instability.
The horizontal line, $\upsilon_{{\rm iso}}=0$,
is the locus of non rotating and/or zero volume
configurations.   The vertical line, $\epsilon
_{31}=1$, is the locus of round$ (a_1=a_2=a_3)$
configurations.   The above mentioned lines
divide the non negative semiplane, $\epsilon_
{31}\ge0$, into four regions (from top left in
clockwise sense): A - oblate-like shapes
with real rotation; B - prolate-like shapes
with real rotation; C - prolate-like shapes
with imaginary rotation; D - oblate-like
shapes with imaginary rotation.   Regions
B and D are forbidden to sequences of
R3 fluids.}
\label{f:syso}
\end{figure}

\begin{figure}
\centering
%\resizebox{\hsize}{!}{\includegraphics{sany.eps}} 
\centerline{\psfig{file=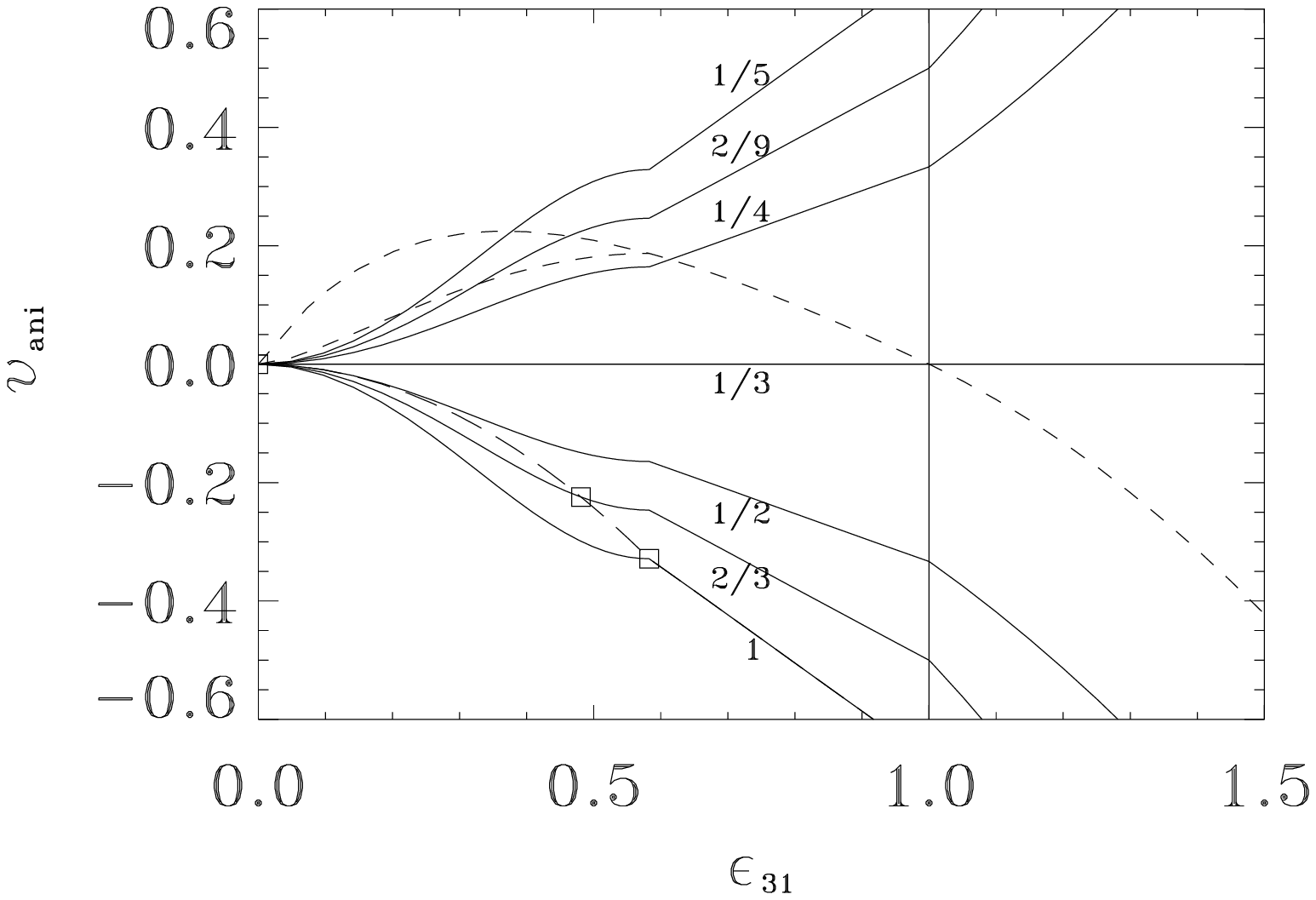,height=130mm,width=90mm}}
\caption{A hypothethical dependence of the
rotation parameter, $\upsilon_{{\rm ani}}$,
related to random motion excess, on the
meridional axis ratio, $\epsilon_
{31}$, with regard to R3 fluids.
Each curve is labelled
by the value of the effective anisotropy
parameter, $\tilde{\zeta}_{33}=\zeta_{33}/
\zeta$.   The horizontal non negative
semiaxis, $\epsilon_{31}\ge0$, $\upsilon_
{{\rm ani}}=0$, is the locus of configurations
with isotropic random velocity distribution,
$\tilde{\zeta}_{33}=1/3$.   The vertical
straight line, $\epsilon_{31}=1$, is the
locus of round $(a_1=a_2=a_3)$ configurations.
The generic sequence starts from a non rotating
configuration (short-dashed lines) and ends at a
configuration where either $\epsilon_{31}=0$
and/or $\tilde{\zeta}_{11}=0$ (long-dashed line).
The regions above the upper short-dashed curve
and below the long-dashed curve, respectively,
are forbidden to R3 fluids.
The non negative vertical semiaxis,
$\upsilon_{{\rm ani}}\ge0$, $\epsilon_{31}=0$,
corresponds to flat $(\tilde{\zeta}_{33}=0)$
configurations with no figure
rotation.   The curves are
symmetric with respect to the horizontal
axis, until the limiting curve, $\tilde{\zeta}_
{33}=\tilde{\zeta}=1$, is reached.
The limiting configuration
where $\tilde{\zeta}_{11}=0$, is marked by a
square: no configuration in virial
equilibrium is allowed on the left,
as it would imply $\tilde{\zeta}_{11}<0$.
No configuration is
allowed on the right of the starting point,
as it would imply $\upsilon<0$.}
\label{f:sany}
\end{figure}

\begin{figure}
\centering
%\resizebox{\hsize}{!}{\includegraphics{s.eps}} 
\centerline{\psfig{file=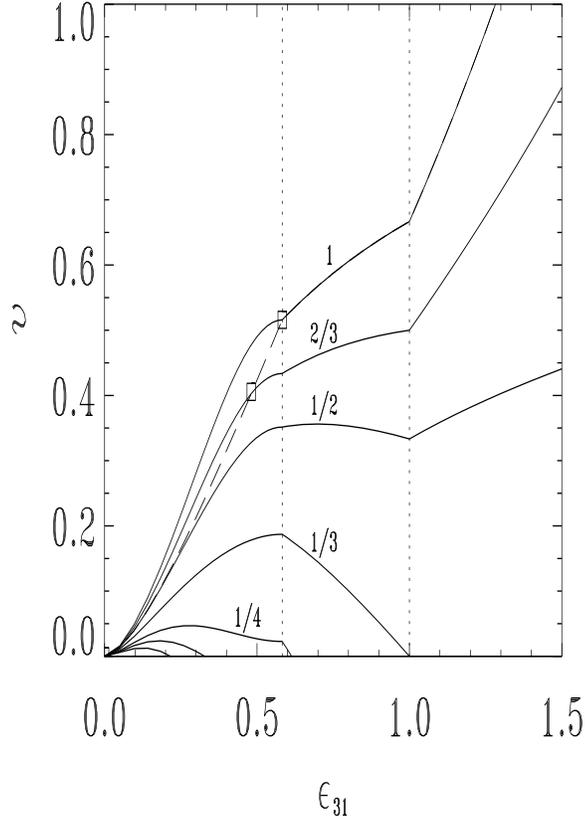,height=130mm,width=90mm}}
\caption{A hypothethical dependence of the
rotation parameter, $\upsilon$,
related to systematic rotation, on the
meridional axis ratio, $\epsilon_
{31}$, with regard to R3 fluids.
Each curve is labelled
by the value of the effective anisotropy
parameter, $\tilde{\zeta}_{33}=\zeta_{33}/
\zeta$, except the lower two, where  
$\tilde{\zeta}_{33}=2/9, 1/5$, respectively.   
The dotted vertical straight lines denote a
hypothetical bifurcation point (left) and
the round $(a_1=a_2=a_3)$ configuration.
The generic sequence starts from a non
rotating configuration on the horisontal
axis and ends at a
configuration where $\epsilon_{31}=0$
or $\tilde{\zeta}_{11}=0$ (long-dashed line),
denoted by a square.   The continuation on the
left of the ending point, where $\tilde{\zeta}
_{11}<0$, $1/2<\tilde{\zeta}_{33}\le1$, is
also shown.   The initial configuration,
related to $\tilde{\zeta}_{33}=1$, corresponds
to $0=a_1=a_2<a_3$ or $a_1=a_2<a_3\to+\infty$,
which is equivalent to $\epsilon_{31}\to+
\infty$.}
\label{f:s}
\end{figure}

Sequences of R3 fluids with assigned
density profiles, can be deduced
from the knowledge of the rotation
parameters, $\upsilon_{{\rm iso}}
(\epsilon_{31})$ and $\upsilon_{{\rm
ani}}(\epsilon_{31},\tilde{\zeta}_
{33})$, as functions of the meridional
axis ratio, $\epsilon_{31}=a_3/a_1$,
and the effective anisotropy parameter,
$\tilde{\zeta}_{33}$, as represented in
Figs.\,\ref{f:syso} and \ref{f:sany},
respectively%
\footnote{Strictly speaking, the
curves plotted in Figs.\,\ref{f:syso},
\ref{f:sany},
and \ref{f:s} are related to the
special case of homeoidally striated
Jacobi ellipsoids (C06), but can be
taken as illustrative for R3 fluids.}.

A hypothetical sequence of axisymmetric
R3 fluids, extended to
the case of imaginary rotation,
is shown in Fig.\,\ref{f:syso}.
Hypothetical dependences of the rotation
parameter, $\upsilon_{{\rm ani}}$, on the
meridional axis ratio, $\epsilon_{31}$,
and the effective anisotropy parameter,
$\tilde{\zeta}_{33}=\zeta_{33}/\zeta$
(labelled on each curve), are shown in
Fig.\,\ref{f:sany}.
With regard to a fixed effective anisotropy
parameter, the related sequence starts from
a nonrotating configuration, 
$\upsilon=0$ i.e. $\upsilon_
{{\rm ani}}=\upsilon_{{\rm iso}}$, and
ends at a configuration where
$\epsilon_{31}=0$ and/or $\tilde{\zeta}_
{11}=0$.   A hypothetical locus of the ending
points related to the latter
alternative, is represented as a
long-dashed curve.   The locus of nonrotating
configurations (short-dashed lines) coincides
with the curves represented in Fig.\,\ref
{f:syso}.   No sequence can be continued on
the right, as imaginary rotation i.e. larger
$\tilde{\zeta}_{33}$ would be
needed and a different sequence
should be considered.   The effect
of positive $(\tilde{\zeta}_{33}<
1/3)$ or negative $(\tilde{\zeta}_
{33}>1/3)$ random motion excess
is equivalent to an additional
real or imaginary rotation,
respectively.
The horizontal non negative
semiaxis, $\epsilon_{31}\ge0$, $\upsilon_
{{\rm ani}}=0$, is the locus of configurations
with isotropic random velocity distribution,
$\tilde{\zeta}_{33}=1/3$.   The vertical
straight line, $\epsilon_{31}=1$, is the
locus of round ($a_1=a_2=a_3$)
%, but not necessarily spherical),
configurations.
                      
With regard to real rotation, the ending
configuration ($\epsilon_{31}=0$ and/or
$\tilde{\zeta}_{11}=0$) is marked by a square.
Configurations on the left are forbidden,
as they would imply negative random
energy tensor component, $(E_{\rm rdm})_{11}
<0$, to satisfy the virial equation (\ref
{eq:viza}), which demands imaginary
rotation around major equatorial axis i.e.
systematic motions other than rotation
around the minor axis.

With regard to imaginary rotation, no
ending point occurs and the system is
allowed to attain negative infinite
rotation parameter, $\upsilon_{{\rm iso}}
\to-\infty$, and infinite meridional
axis ratio, $\epsilon_{31}\to+\infty$.
The related configuration is either a
spindle $(a_1=a_2=0)$ or an infinitely
high cylinder $(a_1=a_2<a_3\to+\infty)$.

Further inspection of Fig.\,\ref{f:sany}
shows additional features, namely: (i)
null left first derivatives on each
sequence at bifurcation point (not
marked therein for sake of clarity), and (ii)
occurrence of symmetric sequences with
respect to the horizontal axis (including
also forbidden configurations).  For
additional considerations on (i), see
C06 (nothing changes with respect to
the special case investigated therein).
The above results may be reduced to a
single statement.
\begin{trivlist}
\item[\hspace\labelsep{\bf Theorem 8.}] \sl
Given a sequence of R3 fluids,
the contribution of random motion excess,
$\upsilon_{{\rm ani}}$,
to the rotation parameter, $\upsilon_
{{\rm iso}}$, has a null left first derivative
at the bifurcation point, $[\diff\upsilon_
{{\rm ani}}/\diff\epsilon_{31}]_{(\epsilon_
{31})_{\rm bif}^-}=0$.
\end{trivlist}

The occurrence of symmetric sequences
with respect to the horizontal axis,
is deduced from Eq.\,(\ref{eq:u3a})
using the condition:
\begin{equation}
\label{eq:sise}
(\tilde{\zeta}_{33}^+)^{-1}-3=-
(\tilde{\zeta}_{33}^-)^{-1}+3~~;
\end{equation}
where $\tilde{\zeta}_{33}^\mp=\zeta_{33}^
\mp/\zeta$ is related to curves
characterized by negative $(\tilde{\zeta}_
{33}=\tilde{\zeta}_{33}^-\ge1/3)$ and
positive $(\tilde{\zeta}_{33}=\tilde
{\zeta}_{33}^+\le1/3)$ values,
respectively, of the rotation parameter,
$\upsilon_{{\rm ani}}$, see Fig.\,\ref
{f:sany}.   Couples of symmetric
sequences (including forbidden
configurations) start from $(\tilde{\zeta}_
{33}^-,\tilde{\zeta}_{33}^+)=(1/3,
1/3)$, where each curve coincides
with the other, and end at (1,1/5).
Sequences in the range, $0\le\tilde
{\zeta}_{33}^+<1/5$, have no symmetric
counterpart.

Let a point, ${\sf P}[\epsilon_{31},
\upsilon_{{\rm iso}}]$, be fixed on
a sequence of axisymmetric R3 fluids,
and the dimensionless moment of inertia
tensor components, ${\cal I}_{11}$ and
${\cal I}_{22}$, be determined by use
of Eqs.\,(\ref{eq:uq3i}) and (\ref
{eq:uqi}).   Let a point, ${\sf P}^
\prime(\epsilon_{31},
\upsilon_{{\rm ani}})$, be fixed on
the plane, $({\sf O}~\epsilon_{31}~
\upsilon_{{\rm ani}})$, and the
effective anisotropy parameters,
$\tilde{\zeta}_{11}$, $\tilde{\zeta}_
{22}$, $\tilde{\zeta}_{33}$, be
determined by use of
Eqs.\,(\ref{eq:uq3a}), (\ref
{eq:u3a}), (\ref{eq:z12L}).  Finally,
let the rotation parameter,
$\upsilon$, be determined by use of
Eq.\,(\ref{eq:uNia}).   Following the
above procedure, sequences of R3
fluids may be generated.   For assigned
density profiles and systematic rotation
velocity fields, there are three
independent parameters, which may
be chosen as two axis ratios,
$\epsilon_{21}$, $\epsilon_{31}$,
and one effective anisotropy
parameter, $\tilde{\zeta}_{33}$.

In the $({\sf O}\epsilon_{31}\upsilon)$
plane (Fig.\,\ref{f:sany}),
each sequence starts from the intersection
between the curves, $\upsilon=[\upsilon
(\epsilon_{31})]_{\rm iso}$ (dashed),
$\upsilon=[\upsilon(\epsilon_{31};
\tilde{\zeta}_{33})]_{\rm ani}$ (full),
and ends at the intersection between
the curves, $\upsilon=\upsilon
(\epsilon_{31}; \tilde{\zeta}_{11}=0)$
(long-dashed), $\upsilon=[\upsilon
(\epsilon_{31}; \tilde{\zeta}_{33})]_
{\rm ani}$ (full).

Hypothetical dependences of the rotation
parameter, $\upsilon$, on the meridional
axis ratio, $\epsilon_{31}$, and the
effective anisotropy parameter, $\tilde
{\zeta}_{33}=\zeta_{33}/\zeta$ (same cases
as in Fig.\,\ref{f:sany}), are shown in
Fig.\,\ref{f:s}.
A hypothetical locus of the ending points
related to $\tilde{\zeta}_{11}=0$ is
represented as a long-dashed curve,
corresponding to $1/2<\tilde{\zeta}_{33}
\le1$.   The continuation of a generic
sequence on the left of the long-dashed
curve would imply $\tilde{\zeta}_{11}<0$
or $\epsilon_{31}<0$.   The ending point
of sequences, related to $0\le\tilde{\zeta}
_{33}<1/2$, coincides with the origin.
The initial configuration, related to
$\tilde{\zeta}_{33}=1$, corresponds to
$0=a_1=a_2<a_3$ or $a_1=a_2<a_3\to+\infty$,
which is equivalent to $\epsilon_{31}\to+
\infty$.

\subsection{A special case: homeoidally striated
Jacobi ellipsoids}\label{profi}

Homeoidally striated Jacobi ellipsoids
are a special case of R3 fluids, for
which the results are already known
(CM05; C06).   Then the particularization
of the current theory to homeoidally
striated Jacobi ellipsoids makes an
useful check.   In the case under
discussion, Eqs.\,(\ref{eq:Prra}) and
(\ref{eq:aref}) reduce to (e.g., CM05;
C06):
\begin{lefteqnarray}
\label{eq:PrrJ}
&& P_{rr}=-\frac{GM^2}{a_1}\nu_{{\rm sel}}
\epsilon_{r2}\epsilon_{r3}A_r~~;\qquad r=1,
2,3~~; \\
\label{eq:ParJ}
&& {\cal P}_{rr}=\left(\frac23\right)^{1/3}
\nu_{{\rm sel}}(\epsilon_{21}\epsilon_{31})^
{1/3}\epsilon_{r2}\epsilon_{r3}A_r~~;\qquad
r=1,2,3~~; \\
\label{eq:aJ}
&& a=\left(\frac23\right)^{1/3}a_1
(\epsilon_{21}\epsilon_{31})^{1/3}~~;
\end{lefteqnarray}
where $\nu_{{\rm sel}}$ is a profile
factor i.e. depending only on the
radial density profile, and $A_r$,
$r=1,2,3,$ are shape factors i.e.
depending on the axis ratios only,
and $a_r$, $r=1,2,3,$ are the
semiaxes of the ellipsoidal boundary.
The dimensionless density profile
may be represented as:
\begin{leftsubeqnarray}
\slabel{eq:profga}
&& \rho=\rho_0f(\xi)~~;\quad f(1)=1~~;\quad\rho_0=
\rho(1)~~; \\
\slabel{eq:profgb}
&& \xi=\frac r{r_0}~~;\quad0\le\xi\le\Xi~~;\quad
\Xi=\frac R{r_0}~~;
\label{seq:profg}
\end{leftsubeqnarray}
where $\rho_0$, $r_0$, are a scaling density and
a scaling radius, respectively, with respect to a
reference isopycnic surface, while $\Xi=\Xi(R,
\theta,\phi)$, and $R$ are related to the truncation
isopycnic surface.

The mass, $M$, and the moment of inertia tensor,
$I_{pq}$, are (e.g., CM05):
\begin{lefteqnarray}
\label{eq:numas}
&& M=\nu_{\rm mas}M_0~~; \\
\label{eq:nuinr}
&& I_{pq}=\delta_{pq}\nu_{\rm inr}Ma_p^2~~;
\end{lefteqnarray}
where $M_0$ is the mass of a homogeneous
ellipsoid with same density and boundary
as the reference isopycnic surface,
and $\nu_{\rm mas}$, $\nu_{\rm inr}$, are profile 
factors.

The combination of Eqs.\,(\ref{eq:I12}),
(\ref{eq:Lrra}), (\ref{eq:aref}), (\ref
{eq:profgb}), and (\ref{eq:nuinr}) yields:
\begin{lefteqnarray}
\label{eq:IqqJ}
&& {\cal I}_{qq}=\frac{I_{qq}}{Ma^2}=
\nu_{\rm inr}\left(\frac23\epsilon_{21}
\epsilon_{31}\right)^{-2/3}\epsilon_{q1}
^2~~; \\
\label{eq:IJ}
&& {\cal I}_3={\cal I}_{11}+{\cal I}_{22}=
\nu_{\rm inr}\left(\frac23\epsilon_{21}
\epsilon_{31}\right)^{-2/3}(1+\epsilon_
{21}^2)~~;
\end{lefteqnarray}
which allow the particularization of the
general results related to R3 fluids, to
homeoidally striated Jacobi ellipsoids.

Using Eqs.\,(\ref{eq:uppq}), (\ref{eq:ParJ}),
(\ref{eq:IqqJ}), and (\ref{eq:IJ}), and
performing some algebra, the rotation
parameter, $\upsilon$, takes the expression:
\begin{equation}
\label{eq:upsh}
\upsilon=\frac23\frac{\nu_{\rm sel}}
{\nu_{\rm inr}}\frac{\zeta_{33}A_q-\zeta_{qq}
\epsilon_{3q}^2A_3}{\zeta_{33}}~~;\qquad q=1,2~~;
\end{equation}
let us define a normalized rotation parameter, as:
\begin{equation}
\label{eq:upzN}
\upsilon_{\rm N}=\frac32\frac
{\nu_{\rm inr}}{\nu_{\rm sel}}\upsilon~~;
\end{equation}
accordingly, Eq.\,(\ref{eq:upsh}) reduces to:
\begin{equation}
\label{eq:vNz}
\upsilon_{\rm N}=\frac{\zeta_{33}A_q-\zeta_{qq}
\epsilon_{3q}^2A_3}{\zeta_{33}}~~;\qquad q=1,2~~;
\end{equation}
which, in spite of a different definition
of the rotation parameter, $\upsilon$,
coincides with a previously known result
(C06) i.e. $\upsilon_{\rm N}=(\upsilon_{\rm N})_{\rm C06}$.

Using Eqs.\,(\ref{eq:zq3}), (\ref{eq:aJ}),
(\ref{eq:IqqJ}), and
performing some algebra, the anisotropy
parameter ratio, $\zeta_{qq}/\zeta_{33}$,
takes the expression:
\begin{equation}
\label{eq:zq3J}
\frac{\zeta_{qq}}{\zeta_{33}}=\epsilon_{q3}^2
\frac{A_q}{A_3}\left(1-\frac{\upsilon_{\rm N}}{A_q}
\right)~~;\qquad q=1,2~~;
\end{equation}
which, keeping in mind a different definition
of the rotation parameter, after some
algebra can be shown to coincide with a
previously known result (C06).   In addition,
Eqs.\,(\ref{eq:zq3J}) disclose that:
\begin{equation}
\label{eq:z1zJ}
\frac{\zeta_{22}}{\zeta_{11}}=\epsilon_{21}^2
\frac{A_2-\upsilon_{\rm N}}{A_1-\upsilon_{\rm N}}~~;
\end{equation}
which, owing to Eqs.\,(\ref{eq:zq3J}),
has necessarily to coincide with a
previously known result (C06).

Using Eqs.\,(\ref{eq:upz}), (\ref{eq:aJ}),
(\ref{eq:IJ}), and
performing some algebra, the alternative
expression of the rotation parameter,
$\upsilon$, takes the form:
\begin{equation}
\label{eq:vNzJ}
\upsilon_{\rm N}=\frac{\zeta_{33}(A_1+\epsilon_{21}^2
A_2+\epsilon_{31}^2A_3)-\zeta\epsilon_{31}^2A_3}
{(1+\epsilon_{21}^2)\zeta_{33}}~~;
\end{equation}
which, in spite of a different definition
of the rotation parameter, $\upsilon$,
coincides with a previously known result
(C06) i.e. $\upsilon_{\rm N}=(\upsilon_{\rm N})_{\rm C06}$.

Using Eqs.\,(\ref{eq:R12}), (\ref{eq:ParJ}),
(\ref{eq:IJ}), and performing some algebra,
the dimensinless moment of inertia tensor component 
ratio, ${\cal I}_{22}/{\cal I}_{11}$, takes
the expression:
\begin{equation}
\label{eq:L21J}
\frac{{\cal I}_{22}}{{\cal I}_{11}}=\epsilon_{21}^2=
\frac{\zeta_{33}\epsilon_{21}^2A_2-\zeta_{22}\epsilon
_{31}^2A_3}{\zeta_{33}A_1-
\zeta_{11}\epsilon_{31}^2A_3}
~~;
\end{equation}
which, after additional algebra, can be
shown to coincide with a previously known
result (C06; the counterpart of ${\cal I}_
{22}/{\cal I}_{11}$ is ${\cal R}_{22}/{\cal
R}_{11}$ therein).

Using Eqs.\,(\ref{eq:up12}), (\ref{eq:ParJ}),
(\ref{eq:IqqJ}), (\ref
{eq:upzN}), and performing some algebra, the
rotation parameter, $\upsilon$, related to
isotropic random velocity distribution
along the equatorial plane, $\zeta_{11}=
\zeta_{22}$, takes the expression:
\begin{equation}
\label{eq:vNiJ}
\upsilon_{\rm N}=\frac{A_1-\epsilon_{21}^2A_2}
{1-\epsilon_{21}^2}~~;
\end{equation}
which, after additional algebra, can be
shown to coincide with a previously known
result (C06).

The above results hold, in particular,
for isotropic random velocity distributions
$(\zeta_{11}=\zeta_{22}=\zeta_{33}=\zeta/3)$,
which implies that the expressions of the
rotation parameters, $\upsilon_{{\rm iso}}$
and $\upsilon_{{\rm ani}}$, via Eq.\,(\ref
{eq:uNo}), must necessarily coincide with
their previously known counterparts (C06).

Using Eqs.\,(\ref{eq:z12L}), (\ref{eq:IqqJ}),
(\ref{eq:IJ}), and performing some algebra,
the anisotropy parameters along the equatorial
plane, $\zeta_{qq}$, $q=1,2,$ take the
expression:
\begin{equation}
\label{eq:zqqJ}
\zeta_{qq}=\frac{\zeta\epsilon_{q1}^2-\zeta_
{33}(2\epsilon_{q1}^2-\epsilon_{2q}^2)}
{1+\epsilon_{21}^2}~~;\qquad q=1,2~~;
\end{equation}
which coincide with previously known
results (C06), in particular for
axisymmetric configurations $(\epsilon_{21}=1)$.

Using Eqs.\,(\ref{seq:z3q0}) and (\ref{eq:L21J}),
the condition for the occurrence of the ending
configuration, related to $\zeta_{11}=0,$ takes
the expression:
\begin{equation}
\label{eq:z10J}
\epsilon_{21}^2=\frac{\zeta_{33}-\zeta_{22}}{\zeta_{33}}=
\frac{2\zeta_{33}-\zeta}{\zeta_{33}}~~;\qquad\zeta_{11}=0~~;
\end{equation}
which coincides with a previously known
result (C06).

Using Eqs.\,(\ref{eq:bife}), (\ref{eq:ParJ}),
(\ref{eq:IqqJ}), and performing some algebra,
a general relation between dimensionless 
self potential-energy tensor components and
dimensionless moment of inertia tensor components,
takes the expression:
\begin{equation}
\label{eq:gecJ}
\frac{\epsilon_{21}^2(A_2-A_1)}{(1-\epsilon_{21}^2)}=
\epsilon_{31}^2A_3~~;
\end{equation}
which coincides with a previously known
result (Caimmi 1996a; C06).   Accordingly,
a necessary condition for the occurrence
of a bifurcation point, Eqs.\,(\ref
{eq:bifc}), reduces to:
\begin{equation}
\label{eq:necJ}
\lim_{\epsilon_{21}\to1}\frac{\epsilon_{21}^2
(A_2-A_1)}{1-\epsilon_{21}^2}=\epsilon_{31}^2
A_3~~;
\end{equation}
which coincides with a previously known
result (Caimmi 1996a; C06).

\section{Conclusion}\label{conc}

The current paper has been aimed in getting more
insight on three main points concerning large-scale
astrophysical systems, namely: (i) formulation of
tensor virial equations from the standpoint of
analytical mechanics; (ii) investigation on the
role of systematic and random motions with respect
to equilibrium configurations; (iii) extent to
which systematic and random motions are equivalent
in flattening or elongating the shape of a mass
distribution.

The tensor virial equations have been
formulated regardless from the nature of the
system and its constituents, by generalizing
and extending a procedure used for the scalar
virial equations, in presence of discrete
subunits (Landau \& Lifchitz 1966, Chap.\,II,
\S\,10).   In particular, the self potential-energy
tensor has been shown to be symmetric with respect
to the indices, $(E_{\rm pot})_{pq}=(E_{\rm pot})_
{qp}$.   Then the results have been extended
to continuous mass distributions.

The role of
systematic and random motions in collisionless,
ideal, self-gravitating fluids, has been analysed in
detail including radial and tangential velocity
dispersion on the equatorial plane, and the
related mean angular velocity, $\overline{\Omega}$,
has been conceived as a figure rotation.

R3 fluids have been
defined as ideal, self-gravitating fluids
in virial equilibrium, with systematic rotation
around a principal axis of inertia.   The
related virial equations have been written in terms
of the moment of inertia tensor, $I_{pq}$,
the self potential-energy tensor, $(E_{\rm pot})_
{pq}$, and the generalized anisotropy tensor,
$\zeta_{pq}$ (CM05; C06).   Additional effort
has been devoted to
the investigation of the properties of
axisymmetric and triaxial configurations.

A unified theory of systematic and random
motions has been developed for R3 fluids, taking
into consideration imaginary rotation 
(Caimmi 1996b; C06).   The effect of
random motion excess has been shown to be 
equivalent to an additional real or
imaginary rotation, inducing flattening (along
the equatorial plane) or elongation (along the
rotation axis), respectively.
Then it has been realized that a R3 fluid
always admits an adjoint configuration
with isotropic random velocity
distribution.

In addition, further constraints have been
established on the amount
of random velocity anisotropy along the 
principal axes, for triaxial configurations.
A necessary condition has been formulated for
the occurrence of bifurcation
points from axisymmetric to triaxial 
configurations in virial equilibrium,
which is independent of the anisotropy
parameters.

The particularization to
the special case of homeoidally striated
Jacobi ellipsoid has been made, and some
previously known results (C06) have been
reproduced.

\section*{Appendix}
\subsection*{A. Tensor potentials and potential-energy tensors}
\label{tepo}

For reasons of simplicity (integrals are easier
than summations to be calculated), let us take
into consideration a continuous distribution of
matter, where volume elements, $\Delta S$,
interact each with the other according to their
charge density, $\rho_\chi=\Delta\phi_\chi/
\Delta S$, being $\Delta\phi_\chi$ the charge
within $\Delta S$.   It is intended that the
results found in this section can be extended
to discrete mass distributions, using summations
instead of integrals.   Let $({\sf O}~x_1~x_2~
x_3)$ be a reference frame where the origin
coincides with the centre of mass, and the
coordinate axes with the principal axes of
inertia.

The effect of the interaction on an infinitesimal
volume element, $\diff^3S=\diff x_1\diff x_2\diff
x_3$, due to a charge distribution of density,
$\rho_\chi(x_1,x_2,x_3)$, is determined by the
potential:
\begin{equation}
\label{eq:Vp}
{\cal V}(x_1,x_2,x_3)=G_\chi\int_S\displayfrac
{\rho_\chi(x_1^\prime,x_2^\prime,x_3^\prime)
\diff^3S^\prime}{\left[\sum_{s=1}^3(x_s-x_s^
\prime)^2\right]^{-\chi/2}}~~;
\end{equation}
where the constants, $\chi$ and $G_\chi$,
specify the nature and the intensity of
the interaction, respectively, and $S$
is the volume filled by the system.

The first derivatives of the potential
with respect to the coordinates, are:
\begin{equation}
\label{eq:dVp}
\frac{\partial{\cal V}}{\partial x_s}=\chi
G_\chi\int_S\displayfrac{\rho_\chi(x_1^\prime,x_2^
\prime,x_3^\prime)(x_s-x_s^\prime)}{\left[
\sum_{s=1}^3(x_s-x_s^\prime)^2\right]^{1-
\chi/2}}\diff^3S^\prime~~;\qquad s=1,2,3~~;
\end{equation}
it can be seen that the functions of the
coordinates,
${\cal V}$ and $x_p\partial{\cal V}/
\partial x_q$, $1\le p\le3$, $1\le q\le3$,
are homogeneous functions of degree $\chi$,
and the Euler theorem holds (e.g., LL66,
Chap.\,IV, \S\,10).

With regard to a selected infinitesimal
volume element, the potential may be thought
of as the tidal energy due to the whole
charge distribution, related to the point
under consideration, with the unit charge
placed therein.   Associated with the
potential, defined by Eqs.\,(\ref{eq:Vp})
and (\ref{eq:dVp}), is the self potential energy:
\begin{equation}
\label{eq:O1}
E_{{\rm pot}}=-\frac12\int_S\rho_\chi(x_1,
x_2,x_3){\cal V}(x_1,x_2,x_3)\diff^3S~~;
\end{equation}
the self potential energy may be thought of as
the tidal energy due to the whole charge
distribution, related to all the
infinitesimal volume elements, provided
any pair is counted only once.

The coincidence of Eqs.\,(\ref{eq:O1})
and (\ref{eq:Vc}) may be verified along
the following steps: (i) write the
alternative expressions of the potential
energy in explicit form, using Eqs.\,(\ref
{eq:Vp}) and (\ref{eq:dVp}); (ii) express
the explicit form of Eq.\,(\ref{eq:Vc})
as a sum of two halves; (iii) with regard
to one half, replace the variables of
integration, $(x_1,x_2,x_3)\leftrightarrow
(x_1^\prime,x_2^\prime,x_3^\prime)$,
keeping in mind that the integrals are
left unchanged; (iv) sum the resulting
two halves and compare with the explicit
form of Eq.\,(\ref{eq:O1}).

To get more information on the charge
distribution, let us define the tensor
potential:
\begin{equation}
\label{eq:Vpqg}
{\cal V}_{pq}(x_1,x_2,x_3)=G_\chi\int_S
\rho_\chi(x_1^\prime,x_2^\prime,x_3^\prime)
\displayfrac{(x_p-x_p^\prime)(x_q-x_q^\prime)}
{\left[\sum_{s=1}^3(x_s-x_s^\prime)^2\right]^
{1-\chi/2}}\diff^3S^\prime~~;
\end{equation}
and the self potential-energy tensor:
\begin{equation}
\label{eq:Opq}
(E_{{\rm pot}})_{pq}=-\frac12\int_S
\rho_\chi(x_1,x_2,x_3){\cal V}_{pq}(x_1,x_2,x_3)
\diff^3S^\prime~~;
\end{equation}
the above mentioned tensors are manifestly
symmetric:
\begin{lefteqnarray}
\label{eq:Vs}
&& {\cal V}_{pq}(x_1,x_2,x_3)={\cal V}_{qp}
(x_1,x_2,x_3)~~; \\
\label{eq:Os}
&& (E_{{\rm pot}})_{pq}=(E_{{\rm pot}})_{qp}~~;
\end{lefteqnarray}
and the related traces equal their scalar
counterparts:
\begin{lefteqnarray}
\label{eq:Vt}
&& \sum_{s=1}^3{\cal V}_{ss}(x_1,x_2,x_3)={\cal V}
(x_1,x_2,x_3)~~; \\
\label{eq:Ot}
&& \sum_{s=1}^3(E_{{\rm pot}})_{ss}=E_{{\rm pot}}~~;
\end{lefteqnarray}
conform to Eqs.\,(\ref{eq:Vp}), (\ref{eq:Vpqg}),
and (\ref{eq:O1}), (\ref{eq:Opq}), respectively.

The coincidence of Eqs.\,(\ref{eq:Opq})
and (\ref{eq:Vpqc}) may be verified along
the following steps: write the
alternative expressions of the
potential-energy tensor in explicit form,
using Eqs.\,(\ref{eq:Vpqg}) and (\ref{eq:dVp});
(ii) express
the explicit form of Eq.\,(\ref{eq:Vpqc})
as a sum of two halves; (iii) with regard
to one half, replace the variables of
integration, $(x_1,x_2,x_3)\leftrightarrow
(x_1^\prime,x_2^\prime,x_3^\prime)$,
keeping in mind that the integrals are
left unchanged; (iv) sum the resulting
two halves and compare with the explicit
form of Eq.\,(\ref{eq:Opq}).

In the special case of gravitational
interaction, $\chi=-1$, $G_\chi=G$
(constant of gravitation),
$\rho_\chi=\rho$ (mass density), the
potential and the potential energy,
both in scalar and in tensor form,
attain their usual expressions known
in literature [C69, Chap.\,2, \S\,10;
see also therein a proof for the
equivalence of Eqs.\,(\ref{eq:Vpqc})
and (\ref{eq:Opq}) - also illustrative
for the equivalence of Eqs.\,(\ref
{eq:Vc}) and (\ref{eq:O1}) - where
the steps outlined above are followed
in a reversed order].

\subsection*{B. An alternative expression of the
generalized anisotropy parameters}
\label{agap}

Let us rewrite Eqs.\,(\ref{eq:viza}) and
(\ref{eq:vizb}) in a more compact notation,
as:
\begin{equation}
\label{eq:vizz}
(1-\delta_{3r})I_{rr}\overline
{\Omega}^2+M\sigma_{rr}^2+(E_{\rm pot})_{rr}=
M\sigma_{rr}^2-M\zeta_{rr}\sigma^2~~;\qquad
r=1,2,3~~;
\end{equation}
and the combination of Eqs.\,(\ref
{eq:virLa}) and (\ref{eq:vizz}) yields:
\begin{equation}
\label{eq:vizIz}
\frac12\ddot{I}_{rr}=M(\sigma_{rr}^2-
\zeta_{rr}\sigma^2)~~;\qquad r=1,2,3~~;
\end{equation}
from which the following expression of the
generalized anisotropy parameters is derived:
\begin{equation}
\label{eq:zitI}
\zeta_{rr}=\frac{\sigma_{rr}^2}{\sigma^2}-
\frac12\frac{\ddot{I}_{rr}}{M\sigma^2}\qquad
r=1,2,3~~;
\end{equation}
and the related trace, owing to Eqs.\,(\ref
{seq:Ipq}), (\ref{eq:vizc}), and (\ref
{eq:vize}) reads:
\begin{equation}
\label{eq:zI}
\zeta=1-\frac12\frac{\ddot{I}}{M\sigma^2}
~~;\qquad r=1,2,3~~;
\end{equation}
where $\zeta$ exceeds unity for a moment
of inertia with respect to the centre of
mass decreasing in time, and vice versa.

\end{document}